\documentclass[10.25pt,twoside,a4paper,openany]{article}
\usepackage[round, sort]{natbib}
\usepackage{setspace}
\usepackage[titletoc,toc]{appendix}
\usepackage{geometry}                
\usepackage[font=small,labelfont=bf]{caption}
\usepackage{hyperref}
\usepackage{xfrac}      
\usepackage[parfill]{parskip}    
\usepackage{graphicx}
\usepackage{amssymb}
\usepackage{epstopdf}
\usepackage{mathtools}
\usepackage{lineno}
\usepackage{rotating}
\usepackage{authblk}
\usepackage{amsmath}
\usepackage{fancyhdr}
\usepackage[usenames, dvipsnames]{color}
\author[1]{\small Joshua D. Carmichael\thanks{Corresponding Author: joshuac@lanl.gov}}
\author[2]{\small Robert Nemzek}
\affil[1]{\footnotesize EES-17, Los Alamos National Laboratory, Los Alamos NM, USA}
\affil[2]{\footnotesize ISR-2, Los Alamos National Laboratory, Los Alamos NM, USA}
\title{Uncertainty in the Predictive Capability of Detectors that Process Waveforms from Explosions}
\begin{document}

\date{}
\maketitle

\label{firstpage}

\begin{abstract}
Explosions near ground generate multiple geophysical waveforms in the radiation-dominated range of their signature fields. Multi-phenomological explosion monitoring (MultiPEM) at these ranges requires the predictive capability to forecast trigger rates of digital detectors that process such waveform data, and thereby accurately anticipate the probability that hypothetical explosions can be identified in operations. To confront this challenge, we derive and compare the predicted and observed performance of three digital detectors that process radio, acoustic and seismic waveform data that record a small, aboveground explosion. We measure this comparison with the peak range in magnitude $\Delta m$ (magnitude discrepancy) over which different performance curves report the same probability of detection $\text{Pr}_{D}$, within an interval of moderate detection probability, and thereby quantify solutions to three topical monitoring questions. In particular, our solutions (1) demonstrate how empirically parameterized detectors that operate in a variable noisy environments provide fair-to-very good forecasting capability to detect small explosions, (2) show that the observed performance of a particular waveform detector can better forecast performance curves constructed from different observations, when compared to theoretical performance curves, and (3) provide an upper bound on detection uncertainty, in terms of a physical source attribute (magnitude).
\end{abstract}
\section*{Introduction: Predictive Detection of Explosion Signatures}
\label{sec:Intro}
Explosions near Earth's surface release multiple waveform and non-waveform signatures. Detonation of explosive charges aboveground coincides with electromagnetic and thermal energy release, emission of chemical particulates, formation of air shocks, and the transmission of this resultant energy into ground as seismic waves. These emissions cumulatively provide data that an explosion occurred, and the remote collection of such data may constitute the only evidence of that explosion. 

Multi-phenomenological explosion monitoring (multiPEM) is a sequence of analysis tasks (detection, association, location, identification) that quantitatively assembles the multiple geophysical signatures of such explosions to better identify and characterize their sources \citep{Carmichael20161}. Whereas focused disciplines study specific emissions from an explosive source (e.g., seismic waves or radionuclides), multiPEM researchers integrate several explosion signatures to provide stronger detection, parameter estimation, or screening capabilities between different sources or processes. For example, earthquake monitoring agencies like the United States Geological Survey (USGS) often exploit seismic and acoustic data to screen near-surface explosions from earthquakes in seismically active mining areas \citep{Pankow20141}. Similarly, nuclear test-ban verification operations monitor radionuclide emissions that originate near locations of seismic explosion sources to more confidently identify nuclear events \citep{Schaff20121}. MultiPEM research thereby promotes the idea that explosions are better characterized when observational evidence of their emissions is assimilated.

Signal detection is the first data processing task required in multi-signature explosion monitoring. This task enables operations to identify the presence of an explosive source by screening statistically significant explosion signatures from background noise. Several geophysical studies document theoretical or semi-empirical assessments of such individual signal detection operations \citep{Carmichael20162,Schaff20081}, but little research exists on detecting multiple signatures triggered by the same source. Fortunately, waveform detectors (versus detectors that process non-waveform data, like particles) described in geophysical literature do share some general traits. Most such detectors process noise contaminated data in multi-sample length processing windows to produce a detection statistic time series. These detectors then evaluate their particular detection statistic to assess two competing hypotheses at each point: that data within the processing window include (1) only non-target signals and/or noise (null hypothesis $\mathcal{H}_{0}$), or (2) that data contain a noisy target signal (alternative hypothesis $\mathcal{H}_{1}$). A detector tests these hypotheses by comparing its statistic against a threshold, which quantifies the significance of the statistic. The amplitude of a target signal that the detector processes, relative to the deviation in background noise, generally determines this significance. If a statistic exceeds its concurrent threshold, it is likely that a noisy signal of sufficient amplitude was processed ($\mathcal{H}_{1}$ true). If the threshold instead exceeds this statistic, it is less likely that a signal of sufficient amplitude was processed ($\mathcal{H}_{0}$ true). 

Signal detection is therefore challenged in monitoring scenarios that target sources which produce explosion waveforms composed of low signal-to-noise ratio (SNR) emissions, so that the distinction between $\mathcal{H}_{1}$ and $\mathcal{H}_{0}$ is unclear. In these scenarios, the predictive capability of signal detector quantifies the accuracy with which that particular detector can accurately forecast its probability of detecting a waveform contaminated by noise.  A high predictive capability implies that researchers can quantify the probability of detecting a signature of a hypothetical explosion, of a given size, on a given day, with high confidence. A low predictive capability means that researchers can forecast a detector's ability to detect an explosion of a prescribed size with only low confidence.

This paper develops and bounds the predictive capability of signal detectors that process waveform signatures with a common metric. To achieve this goal, we estimate the performance of standard detectors that are currently utilized monitoring operation algorithms and consume geophysical waveform data. In particular, we collect radio, acoustic, and seismic emissions following the aboveground detonation of a $\sim10$ kg solid charge chemical explosive with electromagnetic antennae and colocated seismic and acoustic receivers, deployed in the radiation dominated-range of their respective signatures. We then process these data with signature-specific detectors that operate in certain frequency bands and optimize waveform SNR. To quantify the performance of these detectors, we amplitude-scale and infuse template waveforms thousands of times into 12 days of noise records, and then process these data with our detectors. We then compare our predicted detection rates to our observed detection rates to address three practical questions regarding detector predictive capability:

\begin{enumerate}
\item Does the time-averaged, predicted performance $\bar{\text{Pr}}_{D}^{\text{Pre}}$ of a signal detector match its time-averaged, observed performance $\bar{\text{Pr}}_{D}^{\text{Obs}}$? That is, if a detector predictively identifies explosions of magnitude $m$ with probability $\text{Pr}_{D}$, what is the observed, absolute range of magnitudes $\Delta m$ (magnitude discrepancy) that the detector actually identifies explosions, for that same probability?

\item Does the predicted-versus-observed detector performance that is quantified by $\Delta m$ exceed its day-to-day, observed variability? That is, does the predicted performance assembled on day $A$ match observations from day $A$ better than observations assembled on day $B$?

\item What is the range $R \left( \Delta m \right)$ in predicted-versus-observed magnitude discrepancies $\Delta m$? 

\end{enumerate}

To answer these questions, we first describe the theory of signature emission during an aboveground explosion (Section \ref{sec:GroundEmissions}). 
\begin{figure}
\centering
\includegraphics[width=0.85\textwidth]{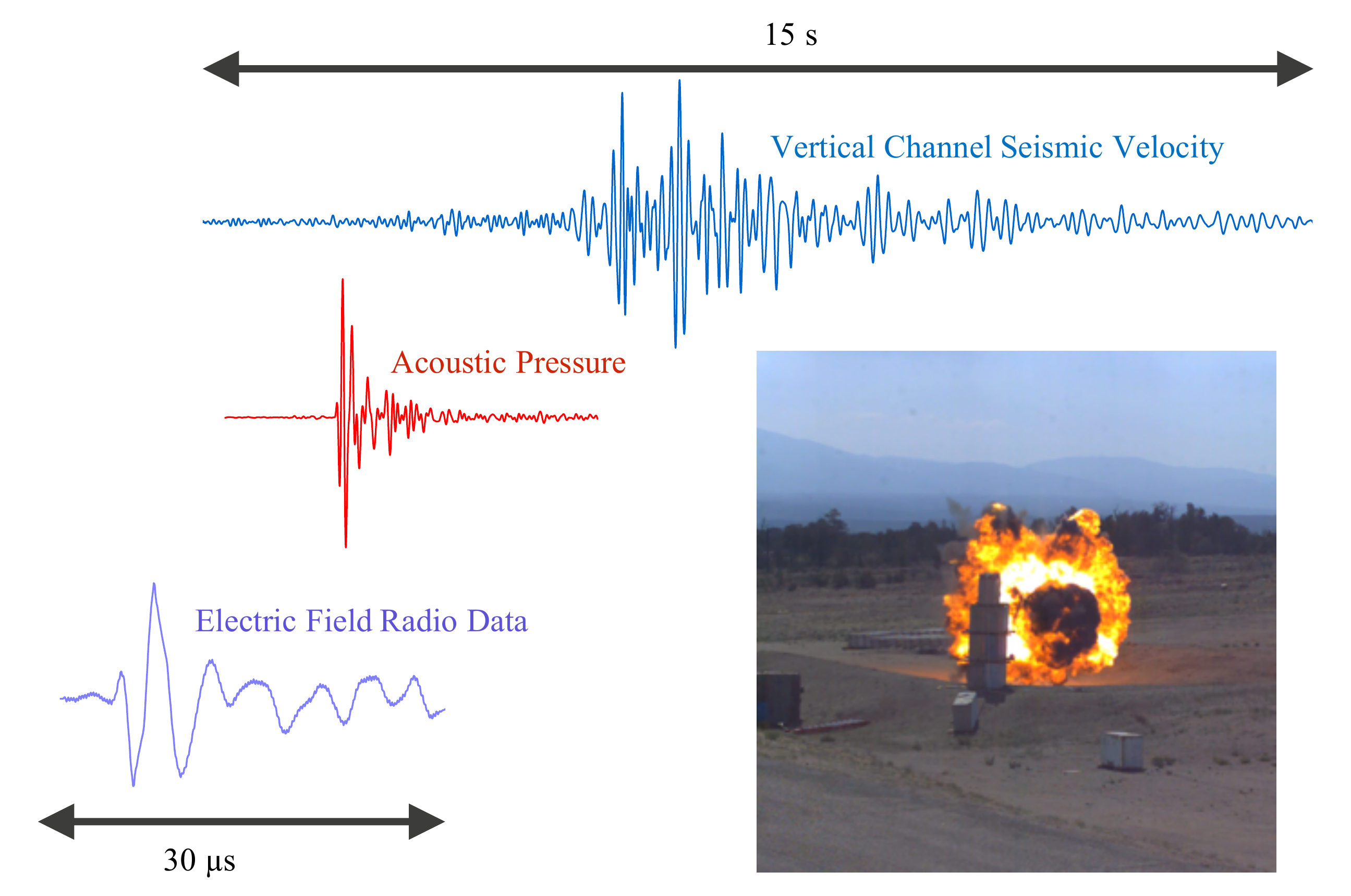}
\caption{Waveforms recorded from an 11.6 kg COMP-B explosive detonated at a 4 m HoB over dry sand, $\sim$ 2 km from source (mechanical waveforms) and $\sim$ 120 m from source (electric waveform).}
 \label{fig:waveforms}
\end{figure}
\section*{Near Ground Explosions Trigger Waveforms}
\label{sec:GroundEmissions}
Explosions produced by bare, solid charges that are detonated near ground trigger multiple waveform signatures (Figure \ref{fig:waveforms}). The combustion reaction that initiates the charges' detonation first triggers a shock within the solid explosive that propagates away from its starting point.  This shock compresses the surrounding air, heating it to temperatures on the order of $10^{4} \text{K}$, sufficient to partially ionize the air. The resulting plasma radiates energy through a mechanism that is still ill-defined. Standard explanations include the creation of a time-varying equivalent dipole due to differential mobility of electrons and ions in the presence of the shock \citep{Harlin20091}, magnetic field compression due to the expanding plasma and subsequent field collapse \citep{Soloviev20021}, and motion of the plasma across the magnetic field. The detonation additionally produces a shock wave in the air that travels many times the speed of sound, and that eventually decays through attenuation and energy absorption into an acoustic wave.

After some time duration following detonation, the air shock no longer heats the air mass to sufficiently high temperature to sustain the radio emissions. However, a second interval of emission starts about that time, that typically lasts longer than initial shock-induced emission interval, and includes more intense radio pulses, indicating a total energy release several times greater than during the air shock interval. For detonations several meters above the ground, this interval still precedes any ground contact, and is temporally coincident with combustion of the byproducts of detonation. Combustion of these byproducts after mixing with air then creates a fireball with temperatures of at least a few thousand degrees. Additionally, the presence of particulates (like soot) in the fireball raises the possibility of additional sources of charging or charge transfer, either through particle-particle collisions or particle-plasma interactions; however, the exact mechanisms by which ionization in the fireball results in radio emission are incompletely known.

Another interval of RF emissions coincides with the arrival of the shock and explosion byproducts at the ground. The reflecting air shock transmits energy into the ground and thereby generates seismic energy directly below the source that thereon travels as a growing, annulus-like load  that is coupled at the air-ground boundary  \citep{Murphy19811}. The shock reflection also re-heats the fireball, providing energy for additional ionization and renewed emission, presumably by physical means very similar to those observed during the initial fireball formation. Even later-time RF emissions that follow the emissions that accompany air-shock reflection appear to involve environment particulates. The onset time should be related to the time required for the shock wave to lift surface particulates (in an above-ground explosion) or to form a crater (in a sub-surface explosion). The dust lifted into the air is probably collisionally (tribo-electrically) charged \citep{Adushkin20041}. Individual charged particulates might generate small electrostatic discharges if they are spatially close enough for their mutual electric field to surpass that of the atmospheric breakdown field. Particulates may also be entrained into a large-scale, circulating structure similar to that seen (on a much larger scale) in a thunderstorm or volcanic eruption. In this case again, the large-scale charge structure might generate fields above the atmospheric breakdown value, resulting in a collective discharge much larger than that from individual grains. The emission from environmental particulates peaks several tens of milliseconds after it starts, but continues at decreasing levels to the extent of current, reference data records (up to 100s of milliseconds).

Near-ground explosions also produce mechanical waveforms as sources of acoustic and seismic energy \citep{Ford20141, Napoli20181}. First, the super-sonic air-shock that engulfs and ionizes air after detonation eventually decays through attenuation and other energy absorption mechanisms into an acoustic wave. This shock also inputs seismic energy to the ground, directly below the explosion, that travels outward from the source nadir point as a growing, annuls-like load coupled at the air-ground boundary \citep{Murphy19811}. The seismic waveforms that record these explosions include low amplitude body waves (p and s waves) that precede slower propagating surface (Rayleigh) waves. The phase velocity of the shallow-layer Rayleigh waves generally decreases with increasing frequency, so that higher frequency Rayleigh waves that propagate near acoustic air speed constructively interfere with the explosion-triggered air wave and build larger amplitude, high frequency Rayleigh waves at far-field ranges ($\sim$ $1-2$ km) from the source \citep{Kitov19971}. Surface seismometers that are deployed at these ranges and that are not isolated from the air then record such signals as a combination of pure ground motion, and a response of the seismometer mass to the transient barometric pressure changes induced by the explosion. Seismic and acoustic scattering off of topographic features and air-density anomalies further contributes to a wave coda that extends the duration and tail of these mechanical signals \citep{Marcillo20141}. 

Noise, environmental background emissions, and ambient anthropogenic activity can superimpose with the waveform records of each of these explosion signatures. Electronic interference has a variety of sources that include lightning (cite), communication-band signals from cell phones and radios, and even emissions of electrostatically discharging particles in dust devils (cite). Acoustic interference and environmental noise sources include wind, cultural activity (Carmichael et al., 2016), thunder, and changes in ambient background atmospheric pressure. These sources of acoustic noise and interference can also couple into the ground as sources of seismic energy. Such background signals further superimpose with any vehicle traffic, tonal noise \citep{Marcillo20181}, and tectonic activity. The variability in these environmental and anthropogenic background emissions, along with the non-stationary noise environment, creates correspondingly variable conditions for recording explosion-triggered waveforms. The amplitude of any such explosion signals, relative to the amplitude of this non-stationary interference and noise, presents the greatest practical challenge to monitoring small, near ground explosions from far-field deployment distances.
\begin{figure}
\centering
\includegraphics[width=0.85\textwidth]{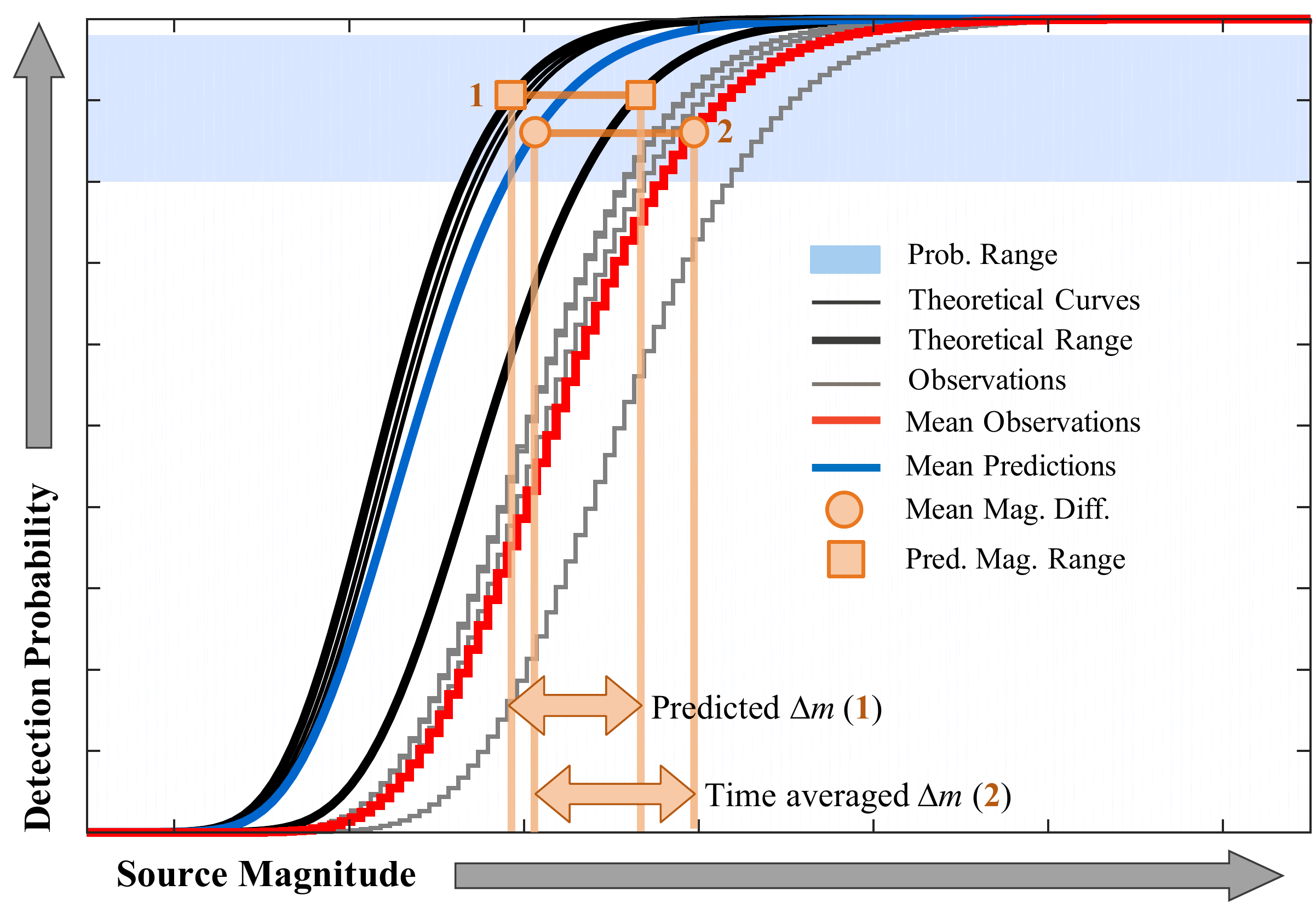}
\caption{ Notational examples of absolute and relative magnitude discrepancy $\Delta m$.  Thin, black curves compare the theoretical performance of a hypothetical signal detector with an $F$-distributed detection statistic against source magnitude. Thicker black curves show their upper and lower performance bounds. The blue curve shows the average of the four ``theoretical'' performance curves $\bar{\text{Pr}}_{D}^{\text{Pre}}$, whereas the red stair-cased curve shows the average $\bar{\text{Pr}}_{D}^{\text{Obs}}$ of the four ``observed'' stair-cased, gray curves. The blue shading indicates a moderate detection probability interval $0.8$ $\le$ $\text{Pr}_{D}$ $\le$ $0.98$. The top-most horizontal line that is terminated by square markers shows the absolute magnitude discrepancy between the range in predicted curves $\text{Pr}^{\text{Pre}}_{D}(m)$ (labeled 1). The lower horizontal line that is terminated by circular markers shows the absolute magnitude discrepancy between $\bar{\text{Pr}}_{D}^{\text{Obs}}$ and  $\bar{\text{Pr}}_{D}^{\text{Pre}}$ (labeled 2). The ratio of the second magnitude discrepancy estimate against the first defines the relative magnitude discrepancy of the time-averaged curves (Equation \ref{eq:magDiscrepRel}).}
 \label{fig:MagDiscrepCartoon}
\end{figure}
\section*{Forecasting Waveform Detection}
\label{sec:detTheoryAssumptions}
To screen an explosion waveform from background noise (radio, acoustic or seismic), we define a signal detector as a decision rule that compares a statistic $s_{k}\left(\boldsymbol{x}\right)$ against a threshold $\eta$ to test if data $\boldsymbol{x}_{k}$ that records signature $k$ is evidence for a target signal (hypothesis $\mathcal{H}_{1}$) or not (hypothesis $\mathcal{H}_{0}$): 
\begin{equation}
s_{k}\left(\boldsymbol{x}\right)  
\,\,
\underset{\mathcal{H}_{0}}
{ \overset{\mathcal{H}_{1}}
{\gtrless}}
\,\,
\eta.
\label{eq:genDecRule}
\end{equation}
The signal detector in Equation \ref{eq:genDecRule} states signature $k$ supports $\mathcal{H}_{1}$ when detection statistic $s_{k}\left(\boldsymbol{x}\right)$ exceeds threshold $\eta$. In practice, this threshold is updated in every data processing window to maintain a constant false alarm on noise probability $\text{Pr}_{FA}$. This means that if $s_{k}\left(\boldsymbol{x}\right)$ has PDF $f_{S}\left(s_{k} \,;\, \mathcal{H}_{0} \right)$ in the absence or sparsity of signal, then $\text{Pr}_{FA}$ relates to $\eta$ through the Neyman Pearson criteria:
\begin{equation}
\begin{split}
\text{Pr}_{FA} &= \int_{\eta}^{\infty} f_{S}\left(\bar{s}_{k} \,;\, \mathcal{H}_{0} \right)  d\bar{s}_{k},
\end{split}
\label{eq:PrFA}
\end{equation}
where $\bar{s}_{k}$ is a dummy integration variable. If $\boldsymbol{x}_{k}$ instead records a target signal produced by a source of magnitude $m$, then $s_{k}\left(\boldsymbol{x}\right)$ has PDF $f_{S}\left(s_{k} \,;\, \mathcal{H}_{1} \right)$. When the absolute source magnitude $m$ is unknown, we refer to a reference magnitude $m_{0}$
and instead parameterize $s_{k}\left(\boldsymbol{x}\right)$ and $f_{S}\left(s_{k} \,;\, \mathcal{H}_{1} \right)$ by relative source magnitude $m-m_{0}$. The theoretical (or predicted) probability $\text{Pr}_{D}^{\text{Pre}}$ of detecting a waveform in $\boldsymbol{x}_{k}$  is then:
\begin{equation}
\begin{split}
\text{Pr}_{D}^{\text{Pre}}(m-m_{0}) &= \int_{\eta}^{\infty} f_{S}\left(\bar{s}_{k} \,;\, \mathcal{H}_{1} \right)  d\bar{s}_{k},
\end{split}
\label{eq:PrD}
\end{equation}
where, again, $\bar{s}_{k}$ is a dummy integration variable. For brevity, we will often write expressions as functions of $m$ rather than $m-m_{0}$, with the understanding that magnitude may be absolute or relative (unitless). Regardless, Equation \ref{eq:PrD} defines the theoretical performance of the signal detector in Equation \ref{eq:genDecRule}. As magnitude $m$ increases, the amplitude of signature $k$ also increases. This increase in signature amplitude decreases the associated overlap between $f_{S}\left(s_{k} \,;\, \mathcal{H}_{0} \right)$ and $f_{S}\left(s_{k} \,;\, \mathcal{H}_{1} \right)$. Such overlap is quantifiable with a scalar noncentrality parameter that is proportional to waveform SNR, so that it is zero under $\mathcal{H}_{0}$, but nonzero and increasing with signal amplitude under $\mathcal{H}_{1}$. This parameter (symbolized as $\lambda$) controls the shape and localization of $f_{S}\left( s_{k} \,;\, \mathcal{H}_{1} \right)$ relative to that of $f_{S}\left(s_{k} \,;\, \mathcal{H}_{0} \right)$. This implies that each noncentrality parameter is itself an increasing function of source magnitude, so that $\lambda$ $=$ $\lambda (m-m_{0})$. We formerly state this dependency as:
\begin{equation}
\label{eq:PrDvLambda}
\text{Pr}_{D}^{\text{Pre}}\left( m_{2} - m_{0} \right) \vert_{\lambda_{2  } } > \text{Pr}_{D}^{\text{Pre}} \left( m_{1}- m_{0}  \right) \vert_{\lambda_{1} }, \quad \text{if:} \,\, \lambda_{2} > \lambda_{1}
\end{equation}
Each waveform detector we implement herein produces a detection statistic $s\left( \boldsymbol{x} \right)$ that we can parameterize by such a noncentrality parameter. Because this parameterization is common between detectors, it provides a means to directly compare their performance. Detector studies conventionally use receiver operating characteristic (ROC) curves to plot $\text{Pr}_{D}(m- m_{0} )$ against false alarm probability $\text{Pr}_{FA}$, for a fixed source attribute (like magnitude $m$). We unconventionally measure detector performance as a comparison between $\text{Pr}_{D}(m- m_{0})$ and relative magnitude $m- m_{0}$, with false alarm probability $\text{Pr}_{FA}$ held constant. Equation \ref{eq:PrD} thereby defines a theoretical performance curve for the detector in Equation \ref{eq:genDecRule} and graphically summarizes whether the detector for signature $k$ is likely to detect an event of magnitude $m$. 

Physical experiments provide data for observed performance curves, to test predictions. Such curves show normalized waveform counts that approximate the theoretical detection probability shown in Equation \ref{eq:PrD} when data statistics are well known. Explicitly, these observed performance curves compare the ratio of detected waveform counts $N_{D}$ to true waveform counts $N_{T}$, versus waveform source magnitude $m- m_{0}$ to define the observed detection probability $\text{Pr}^{\text{Obs}}_{D}$:
\begin{equation}
\label{eq:PrDobs}
\text{Pr}^{\text{Obs}}_{D}(m- m_{0}) = \cfrac{N_{D}\left( m- m_{0} \right)} { N_{T}\left( m- m_{0} \right) }.
\end{equation}
We note that total waveform counts $N_{T}\left( m- m_{0} \right)$ that represent the true number of waveform emissions within a data record is unknown in passive monitoring scenarios. In contrast, $N_{T}\left( m- m_{0} \right)$ is effectively known in certain semi-empirical, signal infusion experiments that we will describe.

We often average many predicted (Equation \ref{eq:PrD}) and observed (Equation \ref{eq:PrDobs}) performance curves that are each associated with an uncertainty or error $\epsilon_{l}$, that is indexed by data processing window or time $l$. In these cases, the weighted mean $\bar{\text{Pr}}_{D}^{\text{Pre}}$ of such performance curves is:
\begin{equation}
\label{eq:meanPrD}
\bar{\text{Pr}}_{D}^{\text{Pre}} =  \cfrac{\displaystyle \sum_{l} \epsilon_{l}^{-1} \text{Pr}_{D}^{\text{Pre}}} { \displaystyle \sum_{l} \epsilon_{l}^{-1}  }
\end{equation}
where $\text{Pr}_{D}^{\text{Pre}}$ is also implicitly indexed by $l$. We omit writing a completely analogous equation for $\bar{\text{Pr}}_{D}^{\text{Obs}}$.
\subsection*{Magnitude Discrepancy Measures Detection Uncertainty}
We define the predictive capability of a detector as the match between the theoretical and observed performance curves over a prescribed probability interval that we bound between lower and upper probability values  ($\text{Pr}_{D}$ $\in$ $[\text{Pr}^{(L)},\,\, \text{Pr}^{(U)}]$). To quantify detection curve agreement (curve match), we measure the maximum range in magnitude $\Delta m$ over which two different performance curves report the same probability of detection $\text{Pr}_{D}$. Conceptually, this absolute ``magnitude discrepancy'' $\Delta m$ shows the uncertainty that a detector can identify waveforms recording an explosion of a given absolute or relative magnitude $m$, when one performance curve serves as a reference for another:
\begin{equation}
\label{eq:absmagDiscrep}
\begin{split}
\text{Pr}_{D}^{(1)} &=  C \quad \text{for:} \,\, m =  m_{1}, \,\, \text{and}
 \\
\text{Pr}_{D}^{(2)} &=  C  \quad \text{for:} \,\, m =  m_{2},  \,\, \text{then:}
 \\
\Delta m &= \underset{ \text{Pr}_{D}  \in [\text{Pr}^{(L)}, \,\, \text{Pr}^{(U)}] } {\max } \vert m_{2} - m_{1}  \vert. 
\end{split}
\end{equation}
Here, $\text{Pr}_{D}^{(1)}$ and $\text{Pr}_{D}^{(2)}$ each indicate distinct detection probability curves, and $C$ is a constant $\text{Pr}^{(L)}$ $<$ $C$ $<$ $\text{Pr}^{(U)}$; again, $m$ is often relative magnitude, in practice. In either case, term $\text{Pr}_{D}^{(1)}$ might present a theoretical detection curve computed for day 1, and $\text{Pr}_{D}^{(2)}$ might present an observed detection curve on day 2; in this example, we would index $\Delta m$ by time to read $\Delta m\left( t_{1}, t_{2} \right)$. We note that because cumulative probability curves monotonically increase, we can represent Equation \ref{eq:absmagDiscrep} with a range over all inverse functions of the form $m^{-1} \left( \text{Pr}_{D} \right)$, but we suggest that this notation is needlessly complicated. Regardless, the conditions listed by Equation \ref{eq:absmagDiscrep} to estimate magnitude discrepancy are equivalent to operations a researcher can perform on graphical plot of performance curves. To execute these graphical operations, we find a horizontal line within the band  $\text{Pr}^{(L)}$ $\le$ $\text{Pr}_{D}$ $\le$ $\text{Pr}^{(U)}$ that (first) intersects curves $\text{Pr}_{D}^{(1)}$ and $\text{Pr}_{D}^{(2)}$, and (second) that maximizes the range $\Delta m$ between their intersection points. We would read this $\Delta m$ value off of the resultant graph as the magnitude discrepancy.

We also define a ``relative'' magnitude discrepancy $\Delta m_{\text{Rel}}$ in terms of the absolute magnitude discrepancy that is given by Equation \ref{eq:absmagDiscrep}. In this relative case, we inversely scale $\Delta m$ by the magnitude range $R^{\,\text{Pre}}\left( m \right)$ between all theoretical detection curves, within that same probability interval:
\begin{equation} 
\label{eq:magDiscrepRel}
\Delta m_{\text{Rel}} =  \cfrac{ \Delta m } { R^{\text{Pre}}\left( m \right) }, \quad \text{where:} \,\, \text{Pr}_{D}  \in [\text{Pr}^{(L)}, \,\, \text{Pr}^{(U)}]
\end{equation} 
We illustrate the utility of this scaling with two thought-examples. In the first case, we consider a family of predicted and observed detection curves that exhibit wide variability over magnitude, but whose average values nearly coincide. In this case, we define $\text{Pr}_{D}^{(1)}$ $=$ $\bar{\text{Pr}}_{D}^{\text{Obs}}$ and $\text{Pr}_{D}^{(2)}$ $=$ $\bar{\text{Pr}}_{D}^{\text{Pre}}$. Both time-averaged curves have little range compared to the range between all $\text{Pr}_{D}^{\text{Pre}}$ curves (since average curves are similar). Hence, the relative magnitude discrepancy $\Delta m_{\text{Rel}}$ is low. In the second case, we consider a family of predicted and observed detection curves that exhibit low intra-curve variability over magnitude. That is, $\text{Pr}_{D}^{\text{Pre}}$ curves and $\text{Pr}_{D}^{\text{Obs}}$ curves are each tightly clustered together over interval $[\text{Pr}^{(L)},\,\, \text{Pr}^{(U)}]$. We further suppose that $\bar{\text{Pr}}_{D}^{\text{Obs}}$ and $\bar{\text{Pr}}_{D}^{\text{Pre}}$ have comparable inter-curve agreement. In this second case, the intra-curve range between all $\text{Pr}_{D}^{\text{Pre}}$ pairs is comparable to the inter-curve range between $\bar{\text{Pr}}_{D}^{\text{Obs}}$ and $\bar{\text{Pr}}_{D}^{\text{Pre}}$ and the relative magnitude discrepancy $\Delta m_{\text{Rel}}$ is higher (Figure \ref{fig:MagDiscrepCartoon}).

In summary, we consider $\Delta m$ to measure the lower bound accuracy of performance curves, $R^{\text{Pre}}\left( m \right)$ to a measure the lower bound predictive precision, and $\Delta m_{\text{Rel}}$ as a precision-normalized measure of accuracy. We then consider time-binned averages of magnitude discrepancy, taken over pairs of performance curves:
\begin{equation}
\label{eq:binDeltam}
\begin{split}
\Delta m\left( t_{1}, t_{2} \right) &= \cfrac{ \displaystyle \sum_{i, j} \cfrac{\Delta m\left( t_{i}, t_{j} \right)}{ \sqrt{ \epsilon_{i} \epsilon_{j}} } } { \displaystyle \sum_{i,j}  \sqrt{  \epsilon_{i}^{-1} \epsilon_{j}^{-1} } }, \,\, \text{where:} \\
t_{1} &\le  t_{i} < t_{j}   \le  t_{2} 
\end{split}
\end{equation}
as an error weighted-estimate of average performance curve accuracy. The error terms $\epsilon_{i}$ and $\epsilon_{j}$ in Equation \ref{eq:binDeltam} are the same error terms from Equation \ref{eq:meanPrD}.
\begin{figure}
\centering
\includegraphics[width=0.85 \textwidth]{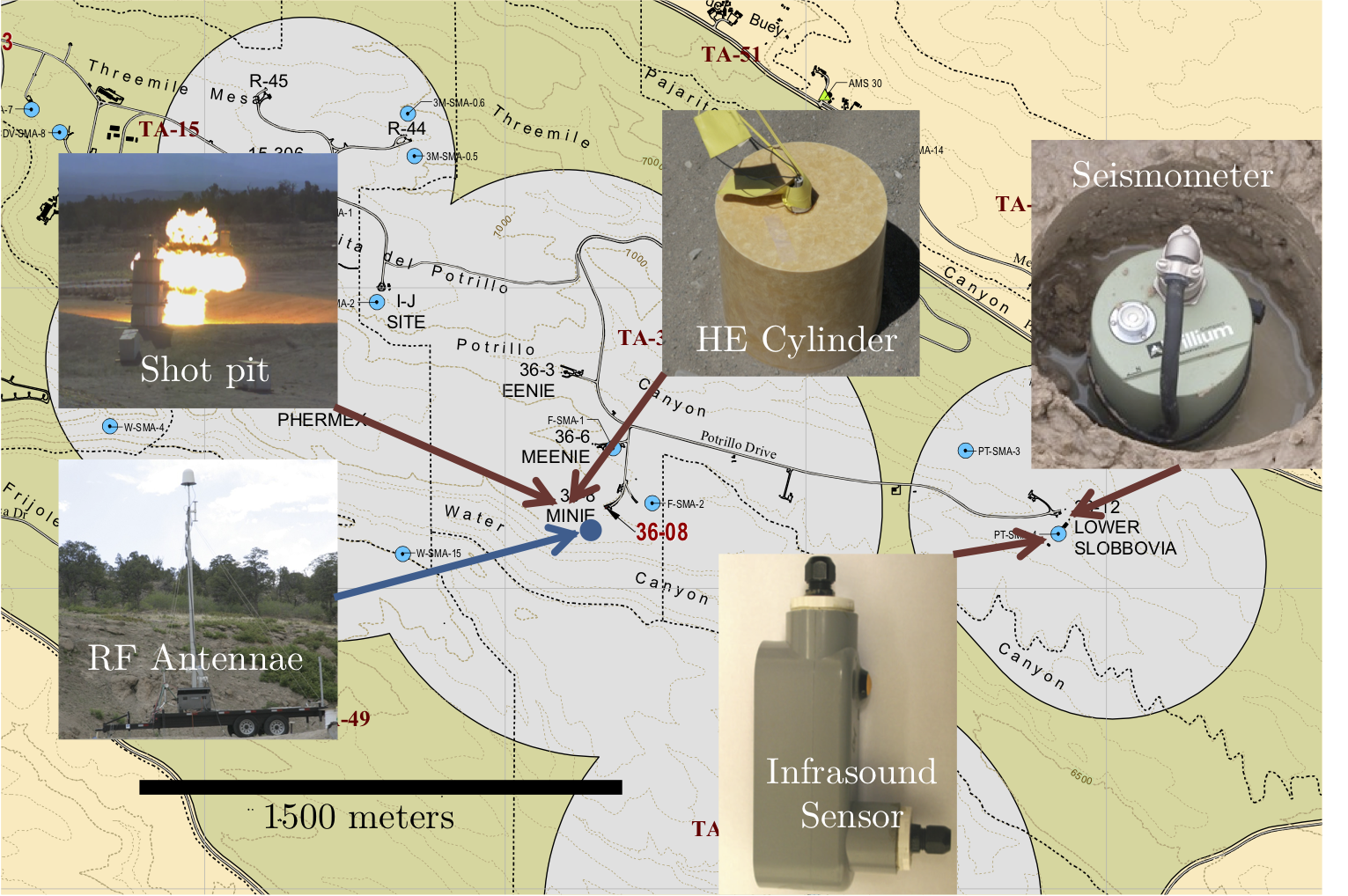}
\caption{Los Alamos National Laboratory material testing ranges. Includes the Minie shot pit, RF antennae, and the seismo-acoustic recording sites.}
 \label{fig:FieldSite}
\end{figure}
\section*{A Multi-Physics Explosion Experiment}
\label{sec:Experiment}
In 2013, Los Alamos National Laboratory conducted an explosive testing campaign using caseless, solid charges of Composition B (COMP-B) detonated below and aboveground at varying distances within the Los Alamos Testing Range (Fig. \ref{fig:FieldSite}).  The charges were deployed at the outdoor, sandy Minie shot pit. Antennae sensors collected hundreds of milliseconds of radio-frequency, electromagnetic emissions preceding and following each charge's detonation. Colocated acoustic and seismic receivers collected continuous, high sample rate ($10^{3}$ s$^{-1}$) mechanical waveform data of the same shots. Electronic clutter included spark plugs firing from a nearby idling firetruck and communication-band signals; however, we filtered some of this clutter out of our analysis band. Acoustic clutter included experiment preparation activity at the shot pit. Long period acoustic noise fluctuations appear driven by changes in background atmospheric pressure. Seismic noise and clutter included these same acoustic sources as well as traffic, and ostensible microseismic activity. The variability in this signal clutter and the non-stationary noise environment created correspondingly variable conditions for the operation of each of these three detectors. We document the specific collection details in each signature specific section below. Previous work on fusion of signatures collected from these same data are further detailed elsewhere \citep{Carmichael20161}.
\subsection*{Explosion Waveform Detection: Summary} \label{app:mechDetect}
We separately compartmentalize the derivation and operational analyses of the radio emission SNR detector (Section \ref{app:Radio}), acoustic emission STA/LTA detector (Appendix \ref{sec:AcousticSTALTA}) and seismic emission correlation detector (Appendix \ref{sec:SeisCorrDetect}). We present each detector as a decision rule on a binary hypothesis test, document the analytical form of the associated detection statistic, estimate the waveform detection threshold and the statistic's unknown parameters, discuss its operation on real data, and derive the detection probability curves. To better compare these curves, we write each PDF under $\mathcal{H}_{1}$ to be shaped by noncentrality parameters $\lambda$ that functionally depend on $m-m_{0}$ (unitless, here). Readers primarily interested in our performance curve comparison can proceed to our Results (Section \ref{sec:analyses}) with little loss in continuity, wherein we cross-reference our detector equations. We further summarize each statistic's arithmetical form (Table \ref{table:detectorStats}), PDF (Table \ref{table:detectorPDFs}), and performance characteristics (Table \ref{table:detectorPerformance}) herein.  We emphasize that our Appendices include analyses of the mechanical waveform detectors.
\begin{table*}[ht]
\caption{Detector Statistic Summaries}
\small
\centering
\begin{tabular}{l | l | l | l | l | l | l | l | l |}
\hline 
Signal, Detector & Detection Statistic &  Estimates  \\  [0.5ex]
\hline 
Radio, SNR & $e\left( \boldsymbol{x} \right) = 10 \cdot  \log_{10} \left[ \cfrac{(N_{}-1) \,\hat{\sigma}_{}^{2}(t > t_{S})} {\sigma^{2}(t < t_{S})  } \right]$   Eq \ref{eq:snrDetStat}  & $\hat{c}, \hat{N}_{E}$  Eq \ref{eq:snrHistNe}
\\  [1ex] 
\hline
Acoustic, STA/LTA & $z \left( \boldsymbol{x} \right) = \cfrac{\hat{\sigma}_{}^{2}(t > t_{S})}{ \hat{\sigma}_{}^{2}(t < t_{S})}$  Eq \ref{eq:acousticVarLta}, \ref{eq:acousticVarSta} &  $\hat{c}, \hat{N}_{1}, \hat{N}_{2}$  Eq \ref{eq:staLtaParamFit}
\\   [1ex] 
\hline
Seismic, Correlation & $r \left( \boldsymbol{x} \right) = 
 \cfrac{ \langle \boldsymbol{ x }, \boldsymbol{u} \rangle_{F}  }{ \vert \vert  \boldsymbol{u}  \vert \vert_{F} \vert \vert   \boldsymbol{x}  \vert \vert_{F}   }$  Eq \ref{eq:decRuleCorrStat}& $\hat{N}_{E}$   Eq \ref{eq:corrEffectDOF}
\\  [1ex]
\hline
\end{tabular}
\label{table:detectorStats}
\end{table*}
\begin{table*}[ht]
\caption{Detector Statistic PDFs}
\small
\centering
\begin{tabular}{l | l | l | l | l | l | l | l | l |}
\hline 
Detection Statistic & Variable that Defines Detector PDF & PDF, Equation\\  [0.5ex]
\hline 
Radio, $e\left( \boldsymbol{x} \right)$ & $\hat{\sigma}_{}^{2}(t > t_{S}) \sim c \chi_{N_{E}}^{2}( \lambda )$ 
$\left\{ \begin{array}{ll}
\mathcal{H}_{0}: \lambda = 0 \\
\mathcal{H}_{1}: \lambda > 0 \\
\end{array} 
\right.$
& $f_{E} \left(  e  ;  \mathcal{H}_{0,1}  \right)$ \quad Eq \ref{eq:dbPdf}
\\  [1ex] 
\hline
Acoustic, $z\left( \boldsymbol{x} \right)$& $z \sim c \mathcal{F}_{N_{1},N_{2}} \left( \lambda  \right)$  
$\left\{ \begin{array}{ll}
\mathcal{H}_{0}: \lambda = 0 \\
\mathcal{H}_{1}: \lambda > 0 \\
\end{array} 
\right.$
& $f_{Z} \left(  z  ;  \mathcal{H}_{0,1}  \right)$  \quad Eq \ref{eq:staLtaFPdf}
\\   [1ex] 
\hline
Seismic, $r\left( \boldsymbol{x} \right)$& $r$ $\sim$ Pearson-Product 
$ \left\{ \begin{array}{ll}
\mathcal{H}_{0}: \lambda = 0 \\
\mathcal{H}_{1}: \lambda > 0 \\
\end{array} 
\right.$
& $f_{R} \left(  r  ;  \mathcal{H}_{0,1}  \right)$ \quad Eq \ref{eq:corrBetaDist}
\\  [1ex]
\hline
\end{tabular}
\label{table:detectorPDFs}
\end{table*}
\begin{table*}[ht]
\caption{Detector Performance}
\small
\centering
\begin{tabular}{l | l | l | l | l | l | l | l | l |}
\hline 
Detector & Pr$_{D}$  & $\hat{\lambda}$ as function of relative magnitude $m-m_{0}$\\  [0.5ex]
\hline
Radio & $1 -  F_{E} \left(  e > \hat{\eta};  \mathcal{H}_{1}  \right)$ \quad Eq  \ref{eq:probDsnr}  &  $10^{2 ( m - m_{0} )} \cfrac{\hat{A}_{0}^{2} N}{\sigma^{2}}$ \quad Eq \ref{eq:radioLambda}
\\ \hline
Acoustic &  $1 -  F_{Z} \left(  z > \hat{\eta} ;  \mathcal{H}_{1}  \right)$  \quad Eq \ref{eq:probDstaLta} &  $10^{2 ( m - m_{0} )}\text{SNR}_{0}^{}(\boldsymbol{x}) \left( \cfrac{\hat{N}_{2}} {\hat{N}_{1}} \right) \left( \hat{N}_{2} - 2 \right) - \hat{N}_{1}$
\quad Eq \ref{eq:staLtaNcEst}\\ \hline
Seismic &  $1 -  F_{R} \left(  r > \hat{\eta};  \mathcal{H}_{1}  \right)$  \quad  Eq \ref{eq:probDcorr} &  $10^{2 (m - m_{0})} \cfrac{\Vert  \boldsymbol{u} \Vert ^{2}}{\hat{\sigma}^{2}}$ \quad Eq \ref{eq:lamCorr}
\\ \hline
\end{tabular}
\label{table:detectorPerformance}
\end{table*}
\subsection*{Radio Emissions and the SNR Detector} \label{app:Radio}
We measured radio emissions with a three-antennae assembly deployed $117\,\text{m}$ from the Minie explosive testing pit. This assembly measured the vertical component of the electric field in high (20 MHz - 1 GHz) and low (2 - 250 MHz) bands, and magnetic field emissions at a $45$-degree orientation to the electric field in a low band (2 - 70 MHz). A capacitive discharge unit triggering system initiated data logging with each shot at detonation time $t_{S}$ and provided up to 500 ms data records, half of which (250 ms) measured pre-detonation background emissions. Digitized electric field data $\boldsymbol{x}$ ($\boldsymbol{x}$  represents a column vector of electric field samples) thereby recorded equal durations of pre-shot ($t < t_{S}$) and post-shot ($t > t_{S}$) fields at two separate sample intervals ($t$ denotes recording time). Our present analyses only include low band electric field measurements and the lower sample rates. To prepare these records for analyses, we first detrended and then bandpass-filtered these data over $20 -150$ MHz to remove communication-band carrier signals using a minimum phase, four-pole Butterworth filter. We then tested these post-processed data for normality and stationarity. First, we performed multiple binning experiments on pre-shot records over several distinct time windows and thereby formed normalized histograms. These histograms fit zero-mean Gaussian curves very well and indicated that the pre-processed radio noise was effectively normal. The presence of explosively-triggered waveform pulses over each post-shot record ($t > t_{S}$) prohibited analogous analyses. Instead, we tested the statistical stationarity of the pre-shot noise to infer its fit for post-shot noise and estimated the true sample variance $\sigma^{2}$ over multiple-duration sliding windows as $\hat{\sigma}^{2}$. These experiments demonstrated that $\hat{\sigma}^{2}$ varied little within each pre-shot record. We therefore assumed that the noise present several ms before each shot also represented background noise present within the shot-triggered emission window. We then used these reference data to estimate the post-shot noise variance. Further, because $\hat{\sigma}^{2}$ varied little, we also considered $\sigma^{2}$ as effectively known. We did \emph{not} assume the number of statistically independent samples $N_{E}$ within data $\boldsymbol{x}$ was known.

\begin{figure}
\centering
\includegraphics[width=\textwidth]{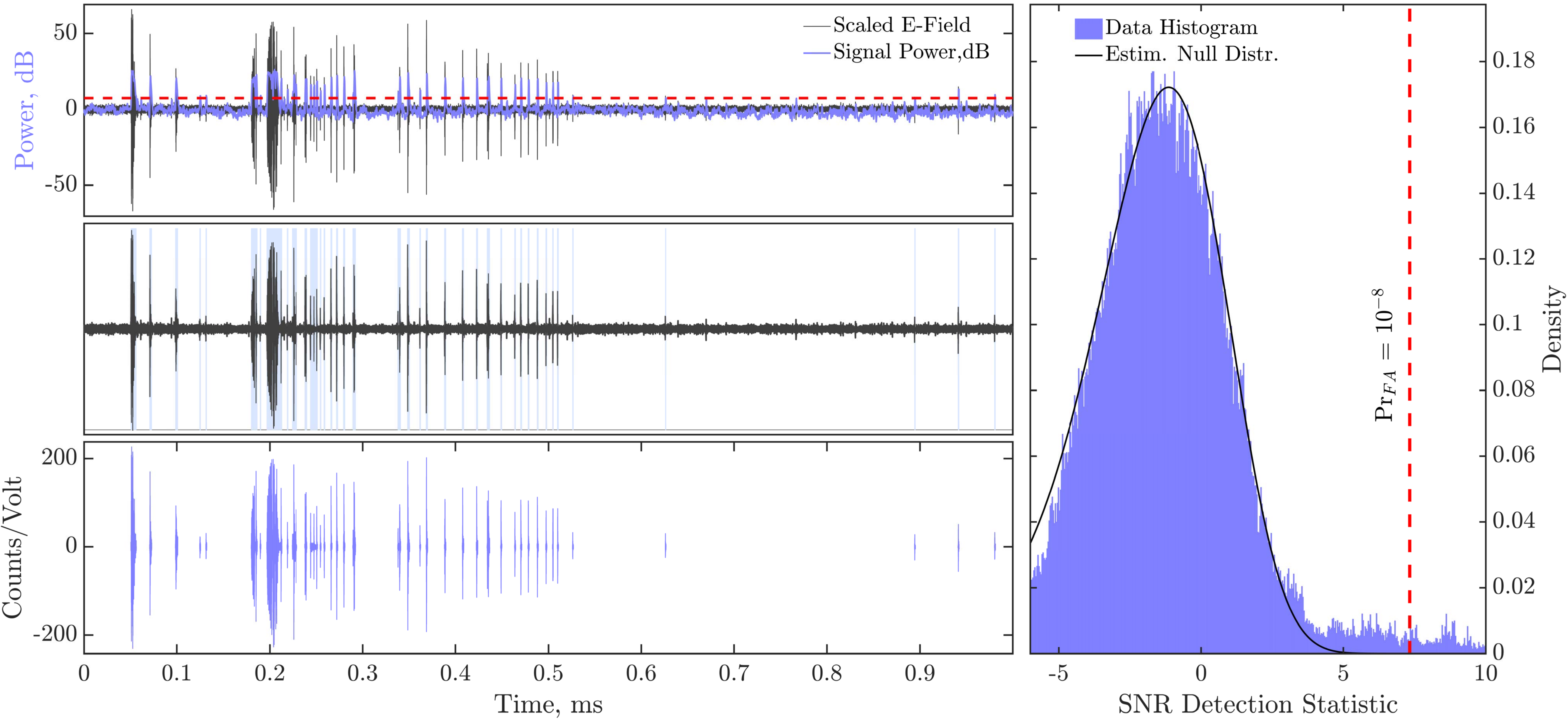}
\caption{ Electric field data $ \boldsymbol{x} $ recorded during over the first $\sim$1 ms proceeding the detonation of an 8' cylindrical COMP-B charge (11.6 kg) at a 4m HoB over dry sand and processed with our SNR detector. \textbf{Left, Top}: Peak-normalized, bandpass filtered (20 -150 MHz) electric waveforms (gray) superimposed with detection statistic $e\left( \boldsymbol{x} \right)$ (purple). The red, horizontal dashed line shows an SNR-threshold for event declaration. \textbf{Left, Middle}: Data from above highlighted blue to show where transient waveform features include sufficient signal to exceed the detector threshold. \textbf{Left, Bottom}: Waveforms extracted from the highlighted data from above. \textbf{Right}: A comparison between the histogram (solid bins) of the detection statistic (Left, Top) and the theoretical null distribution (black curve; Equation \ref{eq:dbPdf}). The vertical dashed line marks the threshold for event declaration consistent with $\text{Pr}_{FA}$ $=$ $10^{-8}$ (Equation \ref{eq:probFAsnr}).}
 \label{fig:RfDetects}
\end{figure}

Having established sufficient stationarity of the noise variance over 250 ms time durations, we identified waveform pulses by processing our antenna data with a noise-adaptive, SNR detector/estimator. Our detector evaluates a binary hypothesis test on the data energy. Under the null hypothesis ($\mathcal{H}_0$), electric field data $\boldsymbol{x}$ includes only noise $\boldsymbol{n}$ throughout the record. Under the alternative hypothesis ($\mathcal{H}_1$), electric field data includes signal $\boldsymbol{u}$ that elevates the data energy over its pre-shot value:
\begin{equation}
\begin{split}
\mathcal{H}_0: &\,\, \sigma^{2}(t > t_{S}) = \sigma^{2}(t < t_{S}), \quad \boldsymbol{x} = \boldsymbol{n}
\\
\mathcal{H}_1: &\,\,  \sigma^{2}(t > t_{S}) > \sigma^{2}(t < t_{S}), \quad \boldsymbol{x} = \boldsymbol{n} +  \boldsymbol{u} 
\end{split}
\end{equation}
Each $N\times1$ column vector ($\boldsymbol{x}$, $\boldsymbol{u}$, and $\boldsymbol{n}$) at sample $l$ contains $l$ through $l+N-1$ consecutive samples of electric field data ($\boldsymbol{x}$), unknown signal ($\boldsymbol{u}$), or Gaussian noise ($\boldsymbol{n}$). The detector estimated post-shot data variance $\sigma^{2}(t > t_{S})$ with a sliding  $N$-point window. This detector thereby computes an $N$-sample energy sum forward from sample $l$ that is defined by:
\begin{equation}
\label{eq:staLtaDetector}
\hat{\sigma}_{l}^{2} = \cfrac{1}{N-1} \displaystyle \sum_{k \,= \, l + 1}^{l + 1 + N} x^{2} \left( k\,\Delta t \right).
\end{equation}
where $x\left( k\,\Delta t \right)$ $=$ $u\left( k\,\Delta t \right)$ $+$ $n\left( k\,\Delta t \right)$ indicates sample $k$ of electric field data $\boldsymbol{x}$, $l$ is the time index, and $\Delta t$ symbolizes the post-decimated sample interval. Deviation of $\hat{\sigma}_{l}^{2}$ from the pre-shot value of $\sigma^{2}(t < t_{S})$ is quantified by a so-called noncentrality parameter $\lambda_{l}$ ($k$ no longer indexes signature type). This scalar is proportional to the SNR of the electric field present within the short-term, $N$-length summation window:
\begin{equation}
\label{eq:snrNc}
\begin{split}
\lambda_{l} &=  \displaystyle \sum_{k \,= \, l + 1}^{l + 1 + N} \cfrac{u^{2} \left( k\,\Delta t \right)}{ \sigma^{2} } 
\\
&= (N-1) \, \text{SNR}^{}\left(\boldsymbol{x} \right), \,\, \text{where:}
\\
\text{SNR}^{}\left(\boldsymbol{x} \right) &= \cfrac{\sum_{k \,= \, l + 1}^{l + 1 + N} u^{2} \left( k\,\Delta t \right)} { N \sigma^{2} }
\end{split}
\end{equation}
We emphasize that $\text{SNR}^{}\left(\boldsymbol{x} \right)$ differs within each $N$-length summation window. With the sample variance and noncentrality parameter defined, the binary hypothesis test is equivalent to:
\begin{equation}
\begin{split}
\mathcal{H}_0: &\,\, \cfrac{(N-1) \, \hat{\sigma}_{l}^{2}(t > t_{S})} {\sigma_{}^{2}(t < t_{S})  } \sim c \chi_{N_{E}}^{2}\left( 0 \right), \quad c, N_{E} \,\, \text{unknown}
\\
\mathcal{H}_1: &\,\,  \cfrac{(N-1) \,\hat{\sigma}_{l}^{2}(t > t_{S})} {\sigma_{}^{2}(t < t_{S})  } \sim c \chi_{N_{E}}^{2}\left( \lambda \right), \quad c, N_{E},\,\lambda \,\, \text{unknown}
\end{split}
\end{equation}
where we omit subscript $l$ on the pre-shot variance $\sigma_{}^{2}(t < t_{S})$ since we assume that it is effectively known and stationary over the pre-shot collection period. Our hypothesis test includes a scaling parameter $c$ that depends on the correlation between neighboring samples that is induced by the presence of structured noise and/or  pre-processing, band-limiting operations. Notationally, $c \chi_{N_{E}}^{2}\left( \lambda \right)$ denotes a scaled noncentral chi-square distribution with $N_{E}$ degrees of freedom, with a noncentrality parameter $\lambda$, and scaling parameter $c$; $N_{E}$ is the effective number of statistically independent samples present in the data processing window that we will discuss estimating later. Increasing values of $\lambda_{l}$ grow more distinct from noise-only variance estimates under $\mathcal{H}_{0}$ and measure the screening power of this test. Our SNR detector measures the scaled variance ratio under each hypothesis in decibels (dB) to form a test statistic $e_{l}\left( \boldsymbol{x} \right)$ at sample $l$ (Figure \ref{fig:RfDetects}, top, purple trace):
\begin{equation}
\label{eq:snrDetStat}
e_{l}\left( \boldsymbol{x} \right) = 10 \log_{10} \left[ \cfrac{(N - 1) \,\hat{\sigma}_{l}^{2}(t > t_{S})} {\sigma^{2}(t < t_{S})  } \right]
\end{equation}
We emphasize that a monotonic (logarithmic) transformation on the variance statistic is unnecessary, since the logarithm's argument is already a sufficient statistic; however, reporting signal power in dB is conventional in radio science. With this convention, the SNR detector compares $e_{l}\left( \boldsymbol{x} \right)$ to an event declaration threshold $\eta_{l}$ (Figure \ref{fig:RfDetects}, red lines) and thereby evaluates the following decision rule:
\begin{equation}
\label{eq:staLtaMaxLike}
e_{l}\left( \boldsymbol{x} \right)
\,
\underset{\mathcal{H}_{0}}
{ \overset{\mathcal{H}_{1}}
{\gtrless}}
\,
\eta_{l}
\end{equation}
We prescribe $\eta_{l}$ using the Neyman Pearson criteria, which fixes a false-alarm on noise rate for the detector. To apply this constraint, we first determine the PDF $f_{E}\left( e; \mathcal{H}_{j} \right)$ for the SNR statistic $e\left( \boldsymbol{x} \right)$ under hypotheses $ \mathcal{H}_{j} $ using a standard variable transformation (sample $l$ omitted hereon in this section). This transformation thereby represents the PDF of the statistic $e$ in terms of the PDF $f_{S}(s;\,\lambda)$ for the noncentral chi-square variable $\chi_{N_{E}}^{2}\left( \lambda \right)$ with noncentrality parameter $\lambda$ (Equation \ref{eq:snrNc}): 
\begin{equation}
f_{E}\left( e; \mathcal{H}_{j} \right) = c \cfrac{1}{10}\ln \left( 10 \right)10^{\frac{1}{10} e} f_{S} \left( c  10^{\frac{1}{10} e}; \,\lambda \right) 
\label{eq:dbPdf}
\end{equation}
Here, $f_{E}\left( e; \mathcal{H}_{0} \right)$ indicates the signal absent, null PDF when $\lambda=0$ and $\mathcal{H}_{0}$ is true. Similarly, $f_{E}\left( e; \mathcal{H}_{1} \right)$ indicates the signal present, alternative PDF when $\lambda>0$ and $\mathcal{H}_{1}$ is true. With these PDFs established, we invert for the value $\eta$ consistent with a right tail probability $\text{Pr}_{FA}$ under $\mathcal{H}_{0}$:
 \begin{equation}
 \label{eq:probFAsnr}
 \begin{split}
\text{Pr}_{FA}  &=  \int_{\eta}^{\infty} f_{E} \left(  e  ;  \mathcal{H}_{0}  \right) dE
\\
&= 1 -  F_{E} \left(  \eta  ;  \mathcal{H}_{0}  \right)
\end{split}
 \end{equation}
where $F_{E} \left( e  ;  \mathcal{H}_{0}  \right)$ is shorthand for the cumulative distribution function (CDF) corresponding to $f_{E} \left(  e  ;  \mathcal{H}_{0}  \right)$. The associated probability that the detector (Equation \ref{eq:staLtaMaxLike}) correctly identifies an electric field pulse buried in noise is the right tail probability under $\mathcal{H}_{1}$,
 \begin{equation}
 \label{eq:probDsnr}
  \begin{split}
\text{Pr}_{D}  &=  \int_{\eta}^{\infty} f_{E} \left(  e  ;  \mathcal{H}_{1}  \right) dE
\\
&= 1 -  F_{E} \left(  \eta ;  \mathcal{H}_{1}  \right),
\end{split}
 \end{equation}
where $F_{E} \left( e ; \mathcal{H}_{1}  \right)$ is the CDF corresponding to $f_{E} \left(  e  ;  \mathcal{H}_{1}  \right)$, as conditioned on $\text{Pr}_{FA}$. The performance of this detector is also conditioned on $c$ and $N_{E}$, which shape both the null and alternative PDFs' curves. Scalar $N_{E}$, in particular, is generally less than it's predicted value of twice the time-bandwidth product of the data. Our detector therefore estimates both $c$ as $\hat{c}$, and $N_{E}$ as $\hat{N}_{E}$ during each data processing operation before setting the detector threshold. To obtain these estimates, we compute normalized histograms of $e\left( \boldsymbol{x} \right)$ using bins spanning the $2.5\%$ to $90\%$ quantiles of $e(\boldsymbol{x})$; this censoring avoids wasteful bin assignment to outlier or identically-zero data. We then numerically estimate $\hat{c}$ and $\hat{N}_{E}$ by minimizing the error between curves of $f_{E} \left(  e  ;  \mathcal{H}_{0}  \right)$ parameterized by $N_{E}$ and the data histogram,
\begin{equation}
\label{eq:snrHistNe}
\hat{c}, \hat{N}_{E} = \underset{c,  N  }{\text{argmax}} \bigr \lvert  \bigr \lvert \text{Hist}\vert_{2.5}^{90}(e) - f_{E} \left(  e ; \mathcal{H}_{0}  \right) \bigr \rvert  \bigr \rvert
\end{equation}
where $\text{Hist}\vert_{2.5}^{90}$ is the histogram operator over the $2.5\%$ to $90\%$ data quantiles. This null-PDF fitting operation assumes that the data are dominated by noise rather than signal, and empirically provides consistent solutions when we initialize $c$ and $N_{E}$ with trial values of one, and twice the time bandwidth product of the post-processed data. We use our resultant estimates $\hat{c}$ and $\hat{N}_{E}$ to then parameterize $f_{E} \left(  e  ;  \mathcal{H}_{0}  \right)$ and calculate detector thresholds $\eta$ as $\hat{\eta}$ (Figure \ref{fig:RfDetects}, vertical line, right). The norm between our quantile-bounded histogram $\text{Hist}\vert_{2.5}^{90}$ and the parameterized PDF  $f_{E} \left(  e  ;  \mathcal{H}_{0}  \right) \vert_{\hat{c}, \hat{N}_{E}}$ measures our fit error $\epsilon$,
\begin{equation}
\label{eq:snrHistError}
\epsilon = \bigr \lvert  \bigr \lvert \text{Hist}\vert_{2.5}^{90}(e) - f_{E} \left(  e ; \mathcal{H}_{0}  \right) \vert_{\hat{c}, \hat{N}_{E}}  \bigr \rvert  \bigr \rvert
\end{equation}
and its square $\epsilon^{2}$ measures fit error variance. We compute processing-window specific detection curves $\text{Pr}_{D}^{\text{Pre}}$ (Equation \ref{eq:probDsnr}) from these best-fit PDFs $f_{E} \left(  e ; \mathcal{H}_{0}  \right) \vert_{\hat{c}, \hat{N}_{E}}$. We compute weighted means $\bar{\text{Pr}}_{D}^{\text{Pre}}$ and $\bar{\text{Pr}}_{D}^{\text{Obs}}$ of such performance curves from Equation \ref{eq:meanPrD}.
\subsection*{Practical Details of Operating the Radio Emission Detector}
\label{sec:detOperation}
In operation, our detector showed that single waveforms often exceeded the SNR statistic's threshold $\eta$ for hundreds to thousands of consecutive, or near consecutive data samples. Figure \ref{fig:RfDetects} illustrates an example of an SNR detector routine that we applied to electric field emissions (top panel) recording an $8$'' (11.6 kg) cylindrical COMP-B charge detonated at a 4 m HoB over dry sand.  Here, SNR estimates exceed the threshold for detection over time durations that are much longer than the sample width of the averaging time window (middle panel, highlighted waveform segments). To avoid redundant triggering on the same signal in these cases, we considered multiple triggers within a prescribed time interval as attributable to the same waveform. Our detector thereby defined the peak SNR value within this time interval as ``the'' detection statistic for the underlying signal. The detector then ignored any subsequent detections within a second time window thereafter to avoid redundant triggering on any coda of the initial, triggering waveform. We set this latter time window to $\sim0.1\mu$s after our manual shot review suggested that shorter waveform features were difficult to interpret due to their low signal complexity. Similarly, if multiple, closely spaced pulse segments were temporally separated over time intervals less than $0.1\mu s$, we defined the larger of the two as a detection on a single, longer $0.1\mu s$ waveform. We then merged these waveform segments into clusters according to pulse duration, regardless of waveform geometry (Figure \ref{fig:RfDetects}, left, bottom panel). To quantify our confidence in the distributional form of these statistics, we estimated errors $\epsilon$ between the predicted null PDF for $e\left( \boldsymbol{x} \right)$ and the data statistics' histogram within each detector operation (Equation \ref{eq:snrHistError}). Figure \ref{fig:RfDetects} (right panel) illustrates an example of misfit between the data histogram and black curve; error $100\% \times \epsilon$ $=$ $12\%$ and instills confidence in our estimation of the signal-absent PDF.
\subsection*{Semi-Empirical and Predicted Radio Emission Performance Curves}
\label{subsec:SNRRocs}
We next assessed the predictive capability of an SNR detector by comparing semi-empirical performance curves $\bar{\text{Pr}}_{D}^{\text{Obs}}$ against theoretically-derived performance curves $\bar{\text{Pr}}_{D}^{\text{Pre}}$ that we parameterized with data-derived statistics. To first construct the empirical performance curves, we selected a 30 $\mu$s electric field waveform that recorded a 11.6 kg charge explosion detonated at a 4 m HoB over dry sand. We then scaled this electric signal's original amplitude $A_{0}$ to prescribed amplitudes $A$ that we sampled from a 100-point SNR grid:
\begin{equation}
\label{eq:magAmpl}
A = 10^{ m - m_{0}} A_{0}
\end{equation}
where $m - m_{0}$ is the magnitude difference between that of the source, which produces the waveform of amplitude $A$, and the magnitude of the source which produces the waveform of amplitude $A_{0}$. We note that neither the absolute magnitude $m_{0}$ of the original source, nor the absolute magnitude $m$ of the source with the scaled waveform need be known to estimate their relative size. This magnitude difference is proportional to the relative SNR of the scaled waveform when measured in decibels (SNR$_{\text{dB}}$):
\begin{equation}
\label{eq:SNRtoMagl}
\begin{split}
\text{SNR}_{\text{dB}} &= 20 \log_{10} \left( \cfrac{A}{A_{0}}\right) 
\\
&= 20 \left( m - m_{0} \right)
\end{split}
\end{equation}
and quantifies event relative magnitude, analogous to that used in seismology (not be confused with magnitude discrepancy $\Delta m$). We do not assume that the source-time function of the scaled explosion source has the same spectral content as the original source, but further consider that such source details are beyond the scope of this paper.

We infused our scaled waveforms hundreds of times into pre-shot noise records collected over 12 days. We repeated this operation over the entire SNR$_{\text{dB}}$ (magnitude) grid for each record, and documented the start time of each infused waveform. Finally, we processed these data with our SNR detector. Each detection routine consumed a data window that contained 68 infused waveforms, produced estimates $\hat{N}_{E}$, and triggered at thresholds $\hat{\eta}$ consistent with a constant false alarm on noise rate (Pr$_{FA}$ $=$ $10^{-8}$). We considered any resultant event declarations to present a true detection if the statistic $e\left( \boldsymbol{x} \right)$ exceeded the threshold $\hat{\eta}$ at a known waveform infusion time, within half of the duration of the short-term processing window. Event declarations outside this window defined false detections. Within our accepted detection windows, we estimated the noncentrality parameter $\lambda$ (Equation \ref{eq:snrNc}) as $\hat{\lambda}$ directly from the detection statistic $e( \boldsymbol{x})$ by combining Equation \ref{eq:snrDetStat} with Equation \ref{eq:snrNc}:
\begin{equation}
\label{eq:radioLambdaHat}
\hat{\lambda} = \hat{N}_{E} 10^{\frac{1}{10} \cdot e}, \,\, \text{where} \,\, e > \hat{\eta}.
\end{equation}
These estimates $\hat{\lambda}$ empirically parameterized our detection count curves. To compare these empirical performance curves against predictions, we used Equation \ref{eq:probDsnr} to compute detection probabilities at each SNR grid point and for each processing day. These probability computations required the true noncentrality parameter $\lambda$ (Equation \ref{eq:snrNc}), additional to the scalars $\hat{N}_{E}$ and $\hat{\eta}$ that we estimated in our initial processing routine (Equation \ref{eq:dbPdf}). We computed these needed $\lambda$ values by exploiting the noncentrality parameter's relation to the square-amplitude estimate $\hat{A}_{0}^{2}$ of the infused signal, as estimated over an $N$-point processing window:
\begin{equation}
\label{eq:radioLambda}
\lambda = 10^{2 ( m - m_{0} ) } \cfrac{\hat{A}_{0}^{2}N}{\sigma^{2}},
\end{equation}
and then computed $\text{Pr}_{D}$ at each SNR (magnitude) grid point. We further scaled these curves by the total number of infused waveforms present in each empirical detection window ($N_{T}\cdot \text{Pr}_{D}$). This parameterization thereby indexed our predicted curves by time and provided consistency in curve comparison.

\begin{figure}
\centering
\includegraphics[width= \textwidth]{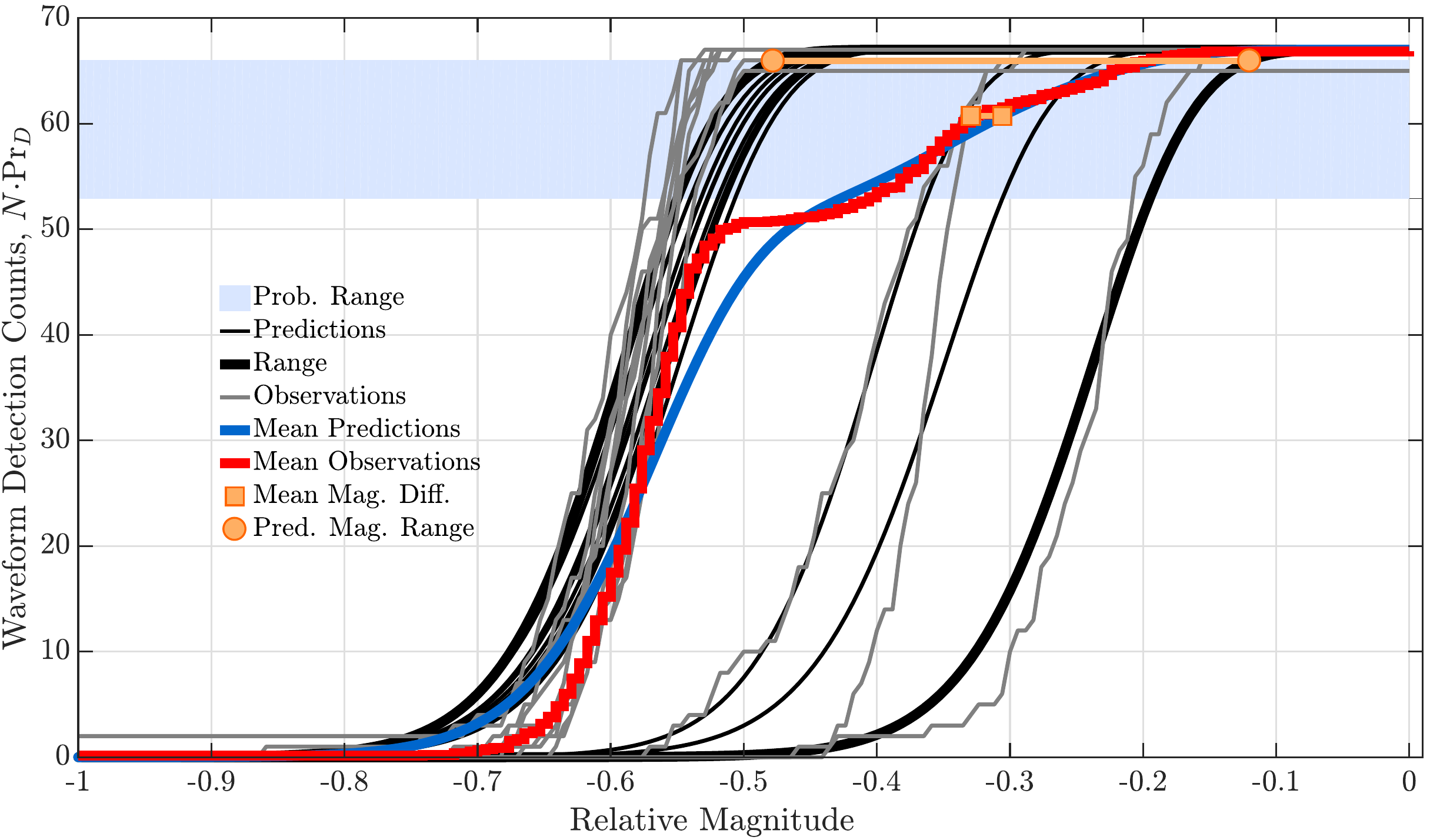}
\caption{ Semi-empirical and theoretical performance curves for the SNR-detector that processed radio emissions measured $117$ m from the Minie Shot pit, displayed as waveform count number versus relative source magnitude.  Data include 68 electric field (V\,m$^{-1}$) waveforms that record 11.6 kg COMP-B charges detonated 4 m over dry ground, which was infused into 12 different days of identically processed noise records. The blue shaded region defines the $0.8$ $\le$ $\text{Pr}_{D}$ $\le$ $0.98$ detection probability range. The red, stair-case curve shows the fit-variance weighted mean (Equation \ref{eq:meanPrD}) of individual detection counts (gray stair-cause curves) versus relative magnitude $m - m_{0}$ for the 12 noise environments, where $m$ is the magnitude of the target, reference event. The smooth blue curves shows the analogous weighted-mean of individual predicted counts (black, smooth curves) in these same 12 noise environments. The horizontal line segment terminated by orange rectangles mark the absolute magnitude discrepancy between the averaged predicted and observed curves within $0.8$ $\le$ $\text{Pr}_{D}$ $\le$ $0.98$. The orange horizontal line segment terminated by circles marks the maximum magnitude range between the detector's theoretical performance. The ratio of these two measures defines the relative magnitude discrepancy $\Delta m_{\text{Rel}}$. These curves are equivalently viewed against $\frac{1}{20}\text{SNR}_{\text{dB}}$ (Equation \ref{eq:SNRtoMagl}).}
 \label{fig:RfRocs}
\end{figure}

Figure \ref{fig:RfRocs} compares the observed performance $\text{Pr}_{D}^{\text{Obs}}$ of the SNR detector against its predicted performance $\text{Pr}_{D}^{\text{Pre}}$ for the 12 different days of processing (displayed as detected waveform counts). The gray staircase plots show the day-specific number of detection counts (minus any false detections) versus relative magnitude $m - m_{0}$ (0 represents unit scaling). The variability in these curves over $m - m_{0}$ reflects the disparate radio noise conditions present each recording day and variability in $\hat{\lambda}$ (Equation \ref{eq:radioLambdaHat}). The solid, smooth curves show the associated, predicted performance curves that are parameterized by the same $\hat{N}_{E}$ and $\hat{\eta}$ values we computed for the empirical processing routines (as well as $\lambda$ from Equation \ref{eq:radioLambda}). The mean predicted curve $\bar{\text{Pr}}_{D}^{\text{Pre}}$ (red staircase plot) and mean observed curve $\bar{\text{Pr}}_{D}^{\text{Obs}}$ (blue solid curve) show our error weighted, 12 day averages (Equation \ref{eq:meanPrD}). We refer the reader to the appendix for analogous analyses of acoustic and seismic detectors.
\section*{Experiment Results}
\label{sec:analyses}
The stated goal of this paper is to quantify the predictive capability of signal detectors that process waveform signatures of aboveground explosions with a common metric. Section \ref{sec:detTheoryAssumptions} introduced the concept of absolute and relative magnitude discrepancy (Equation \ref{eq:absmagDiscrep} and Equation \ref{eq:magDiscrepRel}) to define this metric. We now use this concept to address our three questions (Section \ref{sec:Intro}) that define predictive detection capability, which we summarize and restate here:
\begin{enumerate}
\item Does the time-averaged, predicted performance $\bar{\text{Pr}}_{D}^{\text{Pre}}$ of each detector match its time-averaged, observed performance $\bar{\text{Pr}}_{D}^{\text{Obs}}$? That is, what is the detector's magnitude discrepancy $\Delta m$?
\item Does the predictive capability of a detector exceed it's day-to-day variability? In other words, is the magnitude discrepancy between predicted-versus-observed curves less than the magnitude discrepancy between observed-versus-observed curves?
\item What is the expected range in predicted-versus-observed magnitude discrepancies? 
\end{enumerate}
We sequentially address these questions below for each detector/signature combination.
\subsection*{Radio, SNR}
Figure \ref{fig:RfRocs} compares 12 predicted $\text{Pr}_{D}^{\text{Pre}}$ and observed $\text{Pr}_{D}^{\text{Obs}}$ radio emission, SNR detection curves (Equation \ref{eq:meanPrD}) that are superimposed with their respective, error weighted time-averages, $\bar{\text{Pr}}_{D}^{\text{Pre}}$ and $\bar{\text{Pr}}_{D}^{\text{Obs}}$. The peak range in magnitude between these time-averaged curves where $\bar{\text{Pr}}_{D}^{\text{Obs}}(m)$ $=$ $\bar{\text{Pr}}_{D}^{\text{Pre}}(m)$ $=$ $C$, within the interval of moderate detection probability $0.8$ $\le$ $C$ $\le$ $0.98$ defines the absolute magnitude discrepancy $\Delta m$ $=$ $2.5\times10^{-2}$ (horizontal line segment terminated on red and blue curves with orange squares). To interpret the size of $\Delta m$, we compare it in ratio to the corresponding magnitude range $R^{\text{Pre}}(m)$ over all 12 prediction curves, within the same interval (horizontal line segment terminated with orange circles). This resultant, relative magnitude discrepancy $\Delta m_{\text{Rel}}$ $=$ $6.8 \times10^{-2}$ indicates that the disagreement between the weighted, time averaged curves varies by $\approx7\%$ of our range $R^{\,\text{Pre}}\left( m \right)$ in predictions (Equation \ref{eq:magDiscrepRel}). These data indicate that we can forecast our capability to identify radio emissions from a hypothetical, aboveground explosion that exceeds a certain size, with high confidence, when Equation \ref{eq:staLtaMaxLike} forms our detector. More specifically, we conclude that if explosions produce waveforms $\ge$ $10^{-0.4}$ times smaller in amplitude (or $0.4\times$ smaller, coincidentally) than those recorded from an 11.6 kg at 4m HoB, then our SNR detector will correctly identify such explosion-triggered waveforms with probability $\text{Pr}_{D}$ $\ge$ $0.8$. This agreement means that time-averaged theoretical detection curves represent average observations well. We therefore conclude that $\bar{\text{Pr}}_{D}^{\text{Pre}}$ matches $\bar{\text{Pr}}_{D}^{\text{Obs}}$ for our radio emission, SNR detector, when data record a source $\sim 120$ m away. 

\begin{figure}
\centering
\includegraphics[width=0.85 \textwidth]{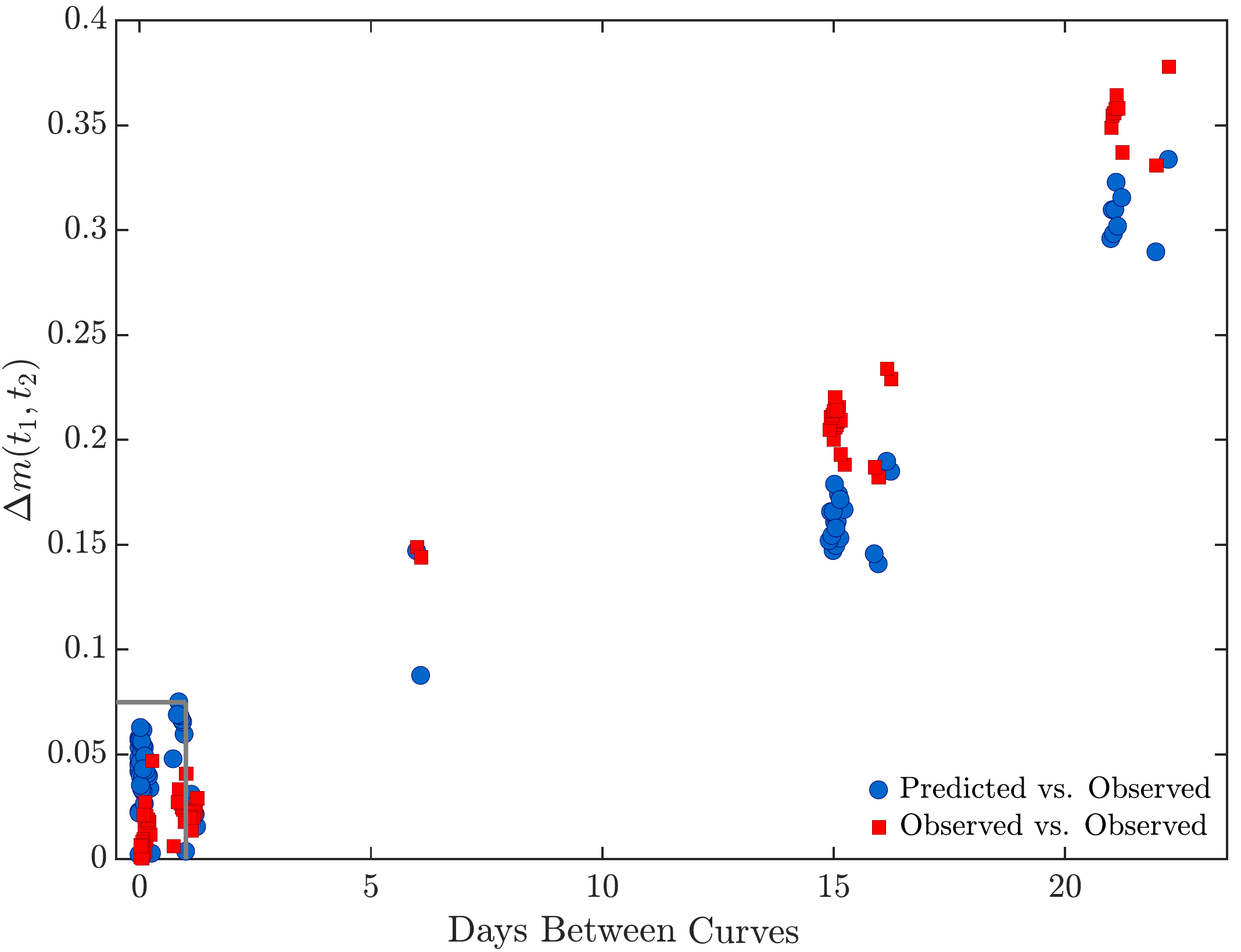}
\caption{ Absolute magnitude discrepancy between performance curves for the radio emission, SNR detector. Each data point shows $\Delta m\left( t_{1}, t_{2}\right)$ between either a pair of predicted-versus-observed performance curves (blue, circular markers), or observed-versus-observed performance curves (red, square markers). The lower left gray box includes discrepancy estimates that compare same-day observations and predictions. We exclude time-averaged performance curves $\bar{\text{Pr}}_{D}^{\text{Pre}}$ and $\bar{\text{Pr}}_{D}^{\text{Obs}}$.}
 \label{fig:RfScatterMagDiscrep}
\end{figure}

\begin{figure}
\centering
\includegraphics[width=0.85 \textwidth]{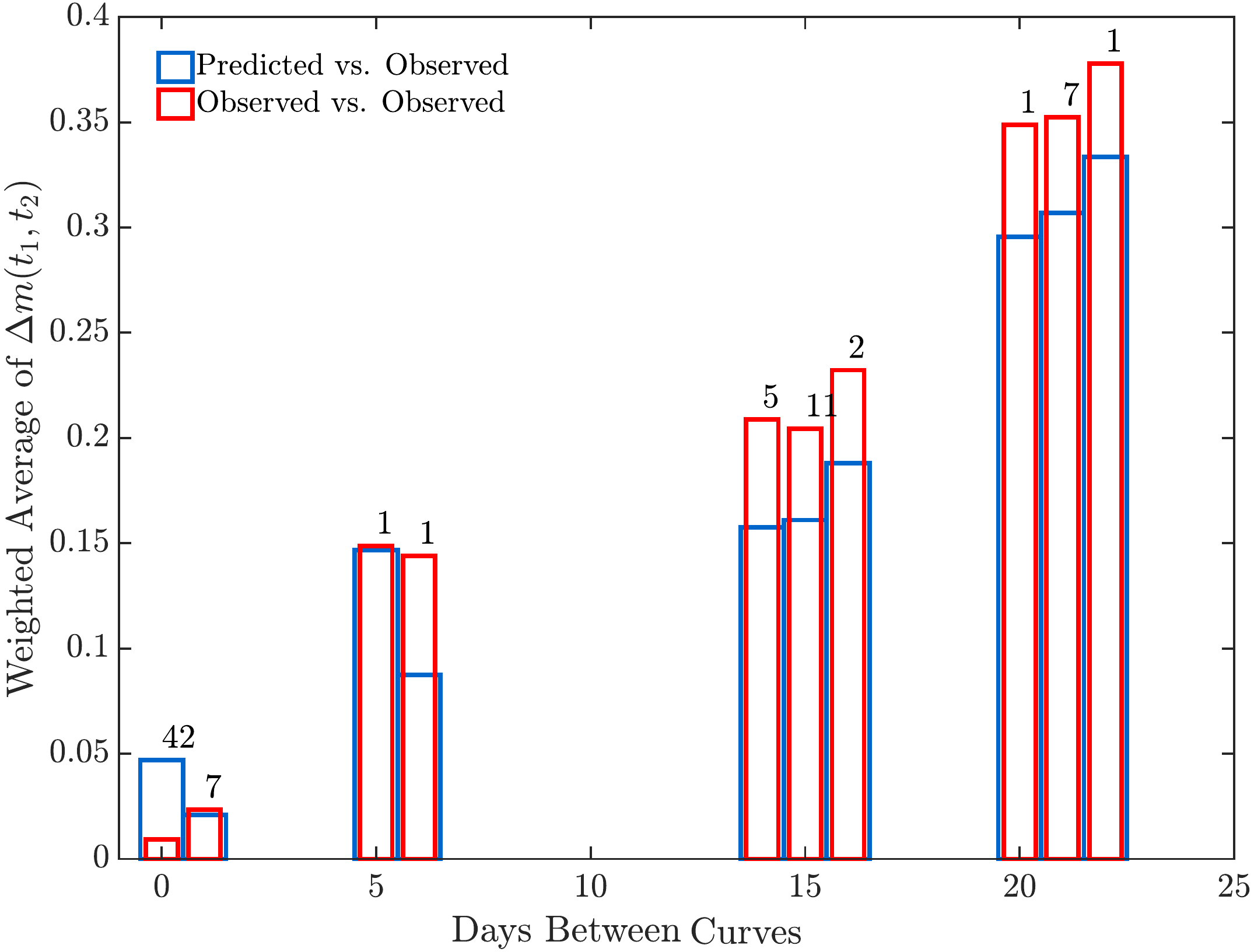}
\caption{ Error weighted estimates of absolute magnitude discrepancy that compare predicted-versus-observed radio emission curves (blue bars), and observed-versus-observed curves (red bars), averaged with Equation \ref{eq:binDeltam}. Numbers above each bar count the discrepancy observations and exclude time-averaged performance curves $\bar{\text{Pr}}_{D}^{\text{Pre}}$ and $\bar{\text{Pr}}_{D}^{\text{Obs}}$.}
 \label{fig:RfBarMagDiscrep}
\end{figure}

In the absence of time-averaged performance curves, either an observed or a predicted performance curve can forecast waveform detection rates for an SNR detector operating on a different day (Figure \ref{fig:RfScatterMagDiscrep}). The magnitude discrepancy between distinct pairs of observed-versus-observed performance curves generally exceeds that between predicted-versus-observed curves (blue markers versus red markers). These data trend upward against differences in observation time and indicate that the predictive capability of the detector correspondingly decreases with time between explosion and forecast of detection capability. Figure \ref{fig:RfBarMagDiscrep} shows that the magnitude discrepancy $\Delta m\left( t_{1}, t_{2}\right)$ between predicted-versus-observed curves is generally less than the magnitude discrepancy between pairs of observed curves, when these data are error weighted and averaged in daily bins (Equation \ref{eq:binDeltam}). This general observation is not uniform. In particular, an observed performance curve can better predict the detection probability of a second, observed detection curve that is calculated on the same day, when compared to a theoretical curve also estimated on that day (leftmost blue versus red bars). This difference is small, however, and exceeded by the relative magnitude discrepancy. We consider this disagreement comparable to average discrepancy and not evidence that theoretical detection curves are less predictive for small time differences. 

Last, we consider the range of magnitude discrepancy values $R\left( \Delta m \right)$ over our 23 days of observation (using a subset of 12 days) to quantify the range in the predictive capability of our radio emission, SNR detector. Referring to Figure \ref{fig:RfBarMagDiscrep}, this range equates to the minimum and maximum heights of the blue bars. We thereby estimate that $2.5\times10^{-2}$ $\le$ $R\left( \Delta m \right)$ $\le$ $3.3\times10^{-1}$. We further normalize $R\left( \Delta m \right)$ over the associated magnitude range $R^{\text{Pre}}\left( m \right)$ $=$ $3.5\times10^{-1}$ of prediction curves in Figure \ref{fig:RfRocs} (orange horizontal line segment terminated by circles). This comparison indicates that the worse-case predictive capability of the SNR detector is comparable to its total range in predicted performance.
\begin{figure}
\centering
\includegraphics[width= \textwidth]{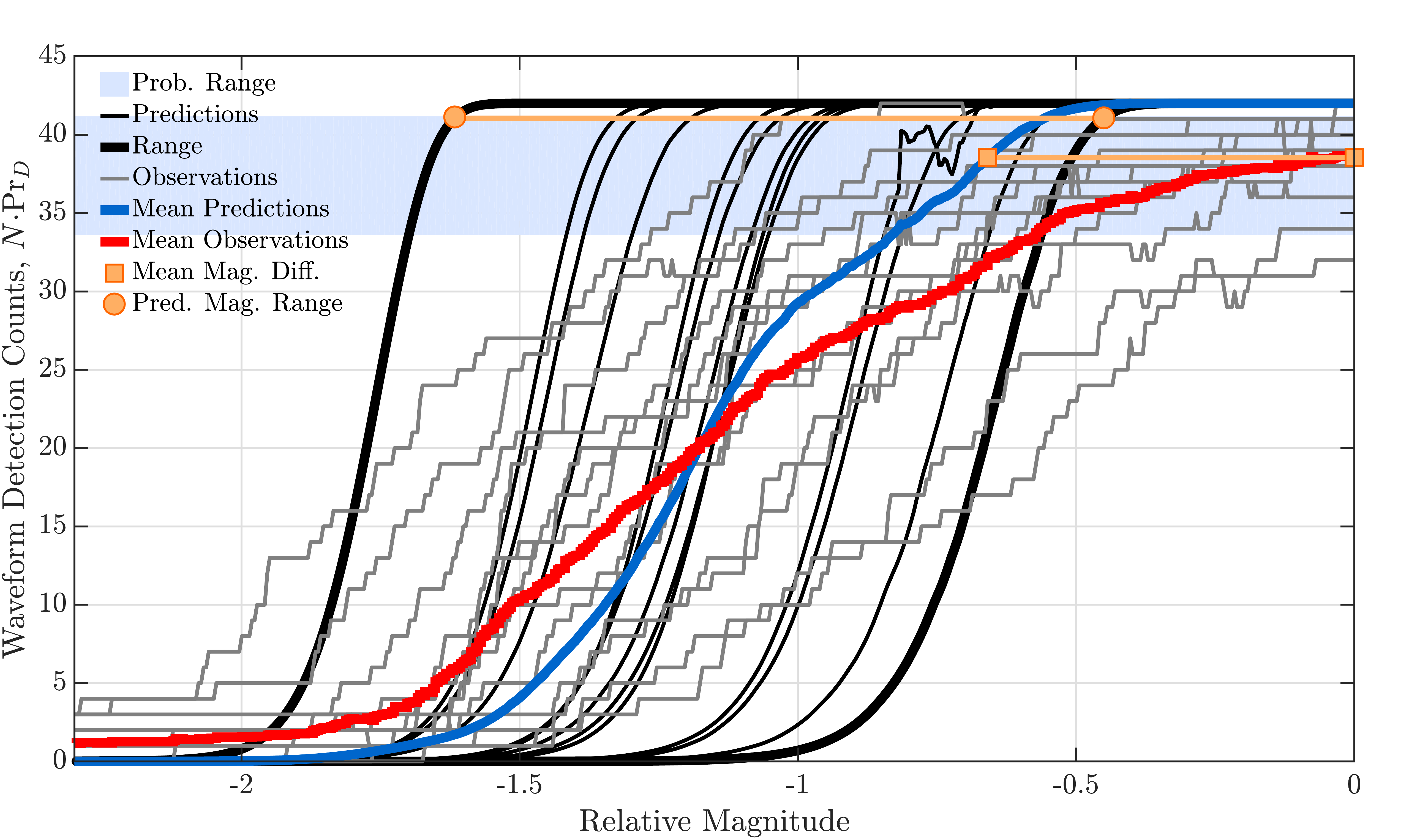}
\caption{ Semi-empirical and theoretical performance curves as shown in Figure \ref{fig:RfRocs}, but for the acoustic emission, STA/LTA-detector. The performance curves exhibit some variability at certain relative magnitudes ($-0.3$, $-0.7$). This variability is driven by erroneous parametric estimates (Equation \ref{eq:staLtaParamFit}) which are themselves driven by large errors $\epsilon$ between data histograms and best-fit central $F$ PDF curves (Equation \ref{eq:staLtaHistError}). Acoustic signals from human activity appeared to cause such errors.}
 \label{fig:AcousticRocs}
\end{figure}
\subsection*{Acoustic, STA/LTA}
Figure \ref{fig:AcousticRocs} compares the 12 observed and predicted performance curves ($\text{Pr}_{D}^{\text{Obs}}$ and $\text{Pr}_{D}^{\text{Pre}}$) for our acoustic emission STA/LTA detector, and each of their error weighted averages ($\bar{\text{Pr}}_{D}^{\text{Obs}}$ and $\bar{\text{Pr}}_{D}^{\text{Pre}}$). The peak magnitude range between the averaged curves within the band of moderate detection probability (shaded region, Figure \ref{fig:AcousticRocs}) gives an absolute magnitude discrepancy $\Delta m$ $=$ $0.66$, where $0$ $\le$ $m-m_{0}$ $\le$ $0.66$. This value is comparable to the magnitude range $R^{\,\text{Pre}}\left( m \right)$ $=$ $1.17$ between predicted curves $\text{Pr}_{D}^{\text{Pre}}$. Their ratio gives a relative magnitude discrepancy of $\Delta m_{\text{Rel}}$ $=$ $5.6 \times10^{-1}$ and indicates that the peak range between time-averaged predicted and observed curves exceeds 50\% the range between all predicted curves. From this value of $\Delta m_{\text{Rel}}$, we conclude that $\bar{\text{Pr}}_{D}^{\text{Pre}}$ shows a poor-to-fair representation for $\bar{\text{Pr}}_{D}^{\text{Obs}}$.  These data indicate that we can forecast our capability to identify acoustic waveforms $\ge$ $10^{-0.6}$ times smaller (or $0.25\times$ smaller) in amplitude that those recorded from an $11.6$ kg charge with only low confidence, when Equation \ref{eq:binDecRuleStaLta} forms our detector, and when the source is $\sim$2 km from the receiver.

Our individual, acoustic emission STA/LTA performance curves are less interpretable. In contrast to the monotonic increase of the radio emission SNR detector performance curves, both the predicted and observed performance curves of the STA/LTA detector show occasional decreases in detected waveform count ($N_{T} \cdot \text{Pr}_{D}$) with increase in relative magnitude $m-m_{0}$. Specifically, periods of alternating low and high waveform detection counts punctuate both curve types, whereby one curve may report $\text{Pr}_{D}\sim 0.9$ at a particular magnitude value, but $\text{Pr}_{D}\le 0.9$ at an incrementally larger  magnitude value. This non-monotonic behavior coincides with large fit errors $\epsilon$ between our data histogram and best estimate for $f_{Z_{}} \left( z; \, \mathcal{H}_0 \right)$ (Equation \ref{eq:staLtaParamFit}). Such errors produced analogously erroneous estimates $\hat{c}$, $\hat{N}_{1}$, and $\hat{N}_{2}$ that inflated or suppressed detection thresholds $\hat{\eta}$, and thereby returned either too few, or far too many detections. During this period, our detector failed to maintain the false alarm constraint $\text{Pr}_{FA}$ that defined $\eta$ (Equation \ref{eq:staLtaFPdf}), as prescribed.

\begin{figure}
\centering
\includegraphics[width=0.85 \textwidth]{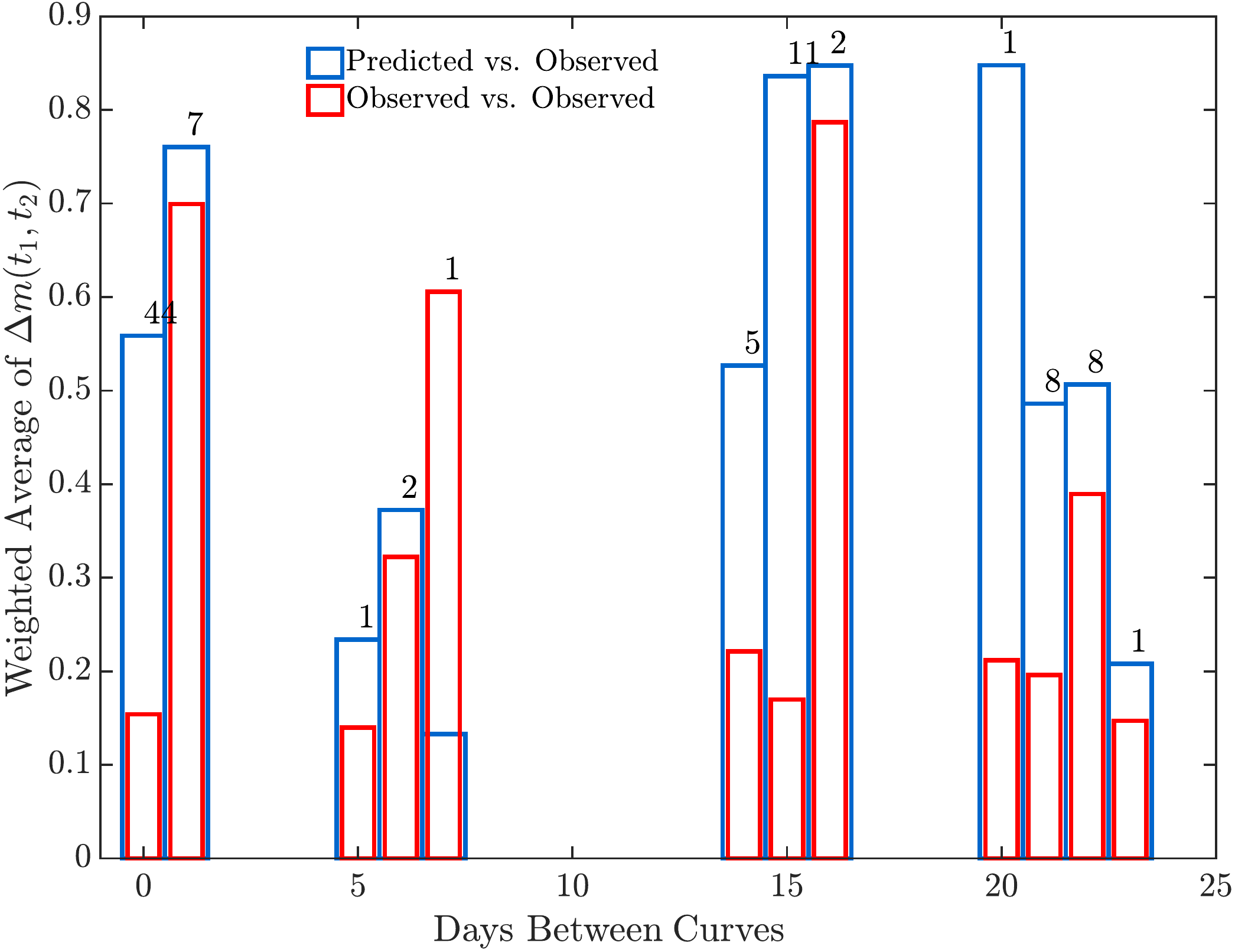}
\caption{ Error weighted averages of the absolute magnitude discrepancy (Equation \ref{eq:binDeltam}) as shown in Figure \ref{fig:RfBarMagDiscrep}, for the acoustic emission STA/LTA detector. Numbers above each bar count discrepancy observations and exclude the time-averaged performance curves $\bar{\text{Pr}}_{D}^{\text{Pre}}$ and $\bar{\text{Pr}}_{D}^{\text{Obs}}$.}
 \label{fig:InfraBarMagDiscrep}
\end{figure}

Our time-binned magnitude discrepancy averages down-weight such errors (Equation \ref{eq:binDeltam}), but still show poorer agreement when compared to the absolute magnitude discrepancy between the time averaged curves (Figure \ref{fig:InfraBarMagDiscrep}). In particular, curve mismatch between predicted and observed performance curves (leftmost blue bars) often exceed observed-versus-observed averages that compare performance curves assembled on different days (rightmost red bars). These data lack a clear trend in magnitude discrepancy $\Delta m\left( t_{1}, t_{2}\right)$ with time  separating each explosion, and our corresponding forecast of explosion waveform detection probability. We conclude that the predictive capability of our STA/LTA detector does not generally exceed it's day-to-day variability, in continued contrast with the performance of the radio emission, SNR detector. 

The overall range of weighted, averaged absolute magnitude discrepancies ($0.16$ $\le$ $R\left( \Delta m \right)$ $\le$ $0.89$) is comparable to the range of magnitudes over which our detector's time-averaged predicted curve $\bar{\text{Pr}}_{D}^{\text{Pre}}$ remains within the interval of moderate detection probability (Figure \ref{fig:AcousticRocs}). That is, $R\left( \Delta m \right)$ is comparable to the magnitude region over which our time-averaged predicted curve decreases from $\bar{\text{Pr}}_{D}^{\text{Pre}}$ $=$ $0.98$, at $m-m_{0}$ $\approx$  $-0.55$, to $\bar{\text{Pr}}_{D}^{\text{Pre}}$ $=$ $0.8$, at $m-m_{0}$ $\approx$ $-0.85$. This means that our daily predicted-versus-observed magnitude discrepancy averages can exceed the entire predicted magnitude range we consider useful for monitoring ($0.8$ $\le$ $\text{Pr}_{D}$ $\le$ $0.98$). We conclude that our acoustic emission STA/LTA detector demonstrates a poor-to-fair capability to forecast true detection rates, when we parameterize our cumulative probability curves with data collected from a single day. Our detector demonstrates an improved (fair, but overly optimistic) capability to forecast time-averaged detection rates that we average from many days of observations (red, stair-cased versus solid, blue curve in Figure \ref{fig:AcousticRocs}).
\begin{figure}
\centering
\includegraphics[width=0.85 \textwidth]{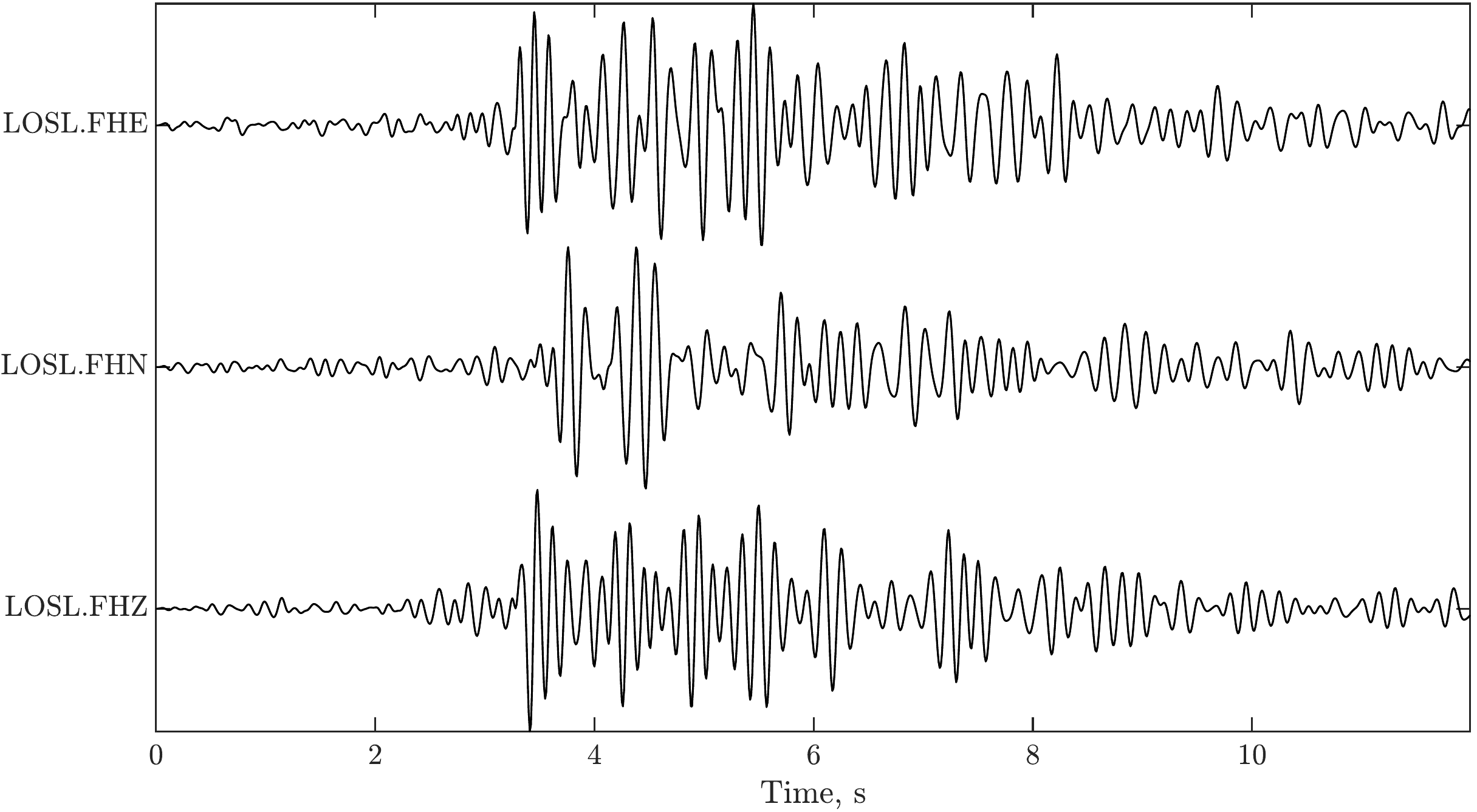}
\caption{ A $12$\,s, three channel waveform template $\boldsymbol{u}$ recording the east (top), north (middle) and vertical (bottom) component of ground motion at station LOSL that was triggered by a 11.6 kg COMP B solid charge detonated at a 4m HoB over dry sand. Data are bandpass filtered to $1.5 - 7.5$Hz with a minimum phase 4-pole Butterworth filter.}
 \label{fig:SeismicTemplate}
\end{figure}
\begin{figure}
\centering
\includegraphics[width= \textwidth]{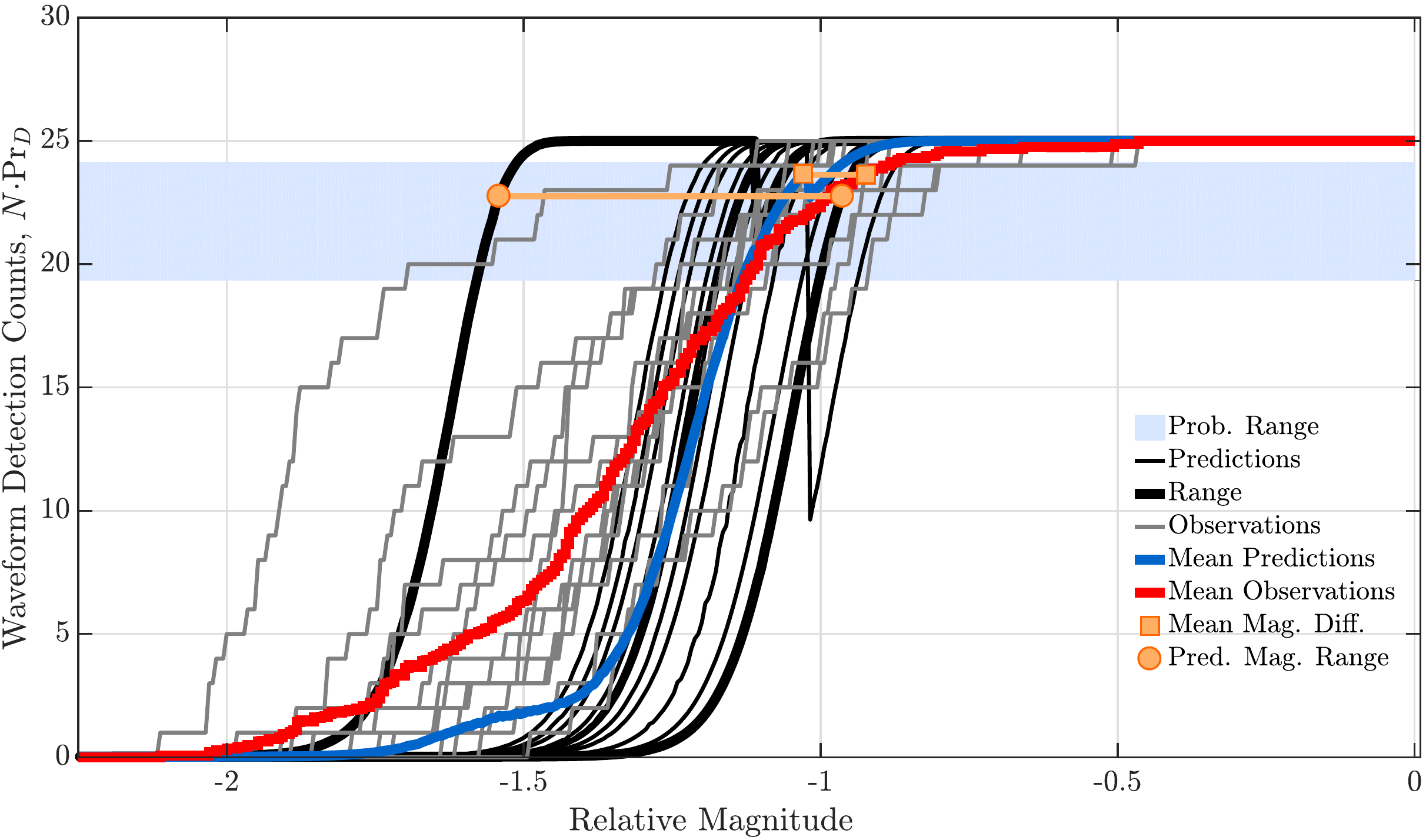}
\caption{ Semi-empirical and theoretical performance curves as shown in Figure \ref{fig:RfRocs}, but for the seismic emission, multi-channel correlation detector. The performance curves exhibit some variability (relative magnitude $\approx$ $-1$) that is driven by erroneous parametric estimates (Equation \ref{eq:corrHistError}), induced here by background seismicity.}
 \label{fig:SeismicRocs}
\end{figure}
\subsection*{Seismic, Correlation}
Figure \ref{fig:SeismicRocs} displays predicted $\text{Pr}_{D}^{\text{Pre}}$, observed $\text{Pr}_{D}^{\text{Obs}}$, and error-weighted average ($\bar{\text{Pr}}_{D}^{\text{Pre}}$, $\bar{\text{Pr}}_{D}^{\text{Obs}}$) performance curves for our seismic emission correlation detector.  The magnitude range between our averaged curves within the band of moderate detection probability  (shaded region, Figure \ref{fig:SeismicRocs}) peaks at an absolute magnitude discrepancy of $\Delta m$ $=$ $0.1$. This value is moderate compared to the magnitude range $R^{\,\text{Pre}}\left( m \right)$ $=$ $0.56$ between predicted curves $\text{Pr}_{D}^{\text{Pre}}$. The absolute magnitude to predicted range ratio gives a relative magnitude discrepancy of $\Delta m_{\text{Rel}}$ $=$ $1.8 \times10^{-1}$ and indicates that the peak range between time-averaged predicted and observed curves is less than 20\% the range between all predicted curves. From this value of $\Delta m_{\text{Rel}}$, we conclude that $\bar{\text{Pr}}_{D}^{\text{Pre}}$ shows a fair-to-good representation for $\bar{\text{Pr}}_{D}^{\text{Obs}}$.  We conclude that we can forecast our capability to identify three-channel seismic waveforms $\ge$ $10^{-1.1}$ times smaller ($0.07\times$ original amplitude) than those recorded from an $11.6$ kg charge with moderate confidence, when Equation \ref{eq:decRuleCorrStat} forms the waveform detector, and when the source is $\sim$2 km from the receiver.

As with our STA/LTA detector, some individual predicted performance curves for the seismic correlation detector decrease with increasing magnitude. In the current case, a predicted performance curve is punctuated by a discontinuous change in detected waveform counts ($N_{T} \cdot \text{Pr}_{D}$), whereby the rightmost curve indicates that our detector regains some capability near magnitude $m-m_{0}$ $=$ $1$, when compared to incrementally larger magnitude values. This behavior coincides with a large fit error $\epsilon$ between our data histogram and our best estimate for $f_{R} \left( r; \, \mathcal{H}_0 \right)$ (Equation \ref{eq:corrBetaDist}). This error produces a poor estimate $\hat{N}_{E}$ for $N_{E}$ (Equation \ref{eq:corrEffectDOF}) where data appear correlated over temporal durations comparable to one-quarter of the detector template length (not shown). This poor estimate also suppresses the detector threshold $\hat{\eta}$, and inflates waveform detection rates. Consequently, the false alarm constraint that defined $\eta$ is inapplicable, and exceeds $\text{Pr}_{FA}$ $=$ $10^{-8}$, as with the acoustic emission, STA/LTA detector.

\begin{figure}
\centering
\includegraphics[width= 0.85\textwidth]{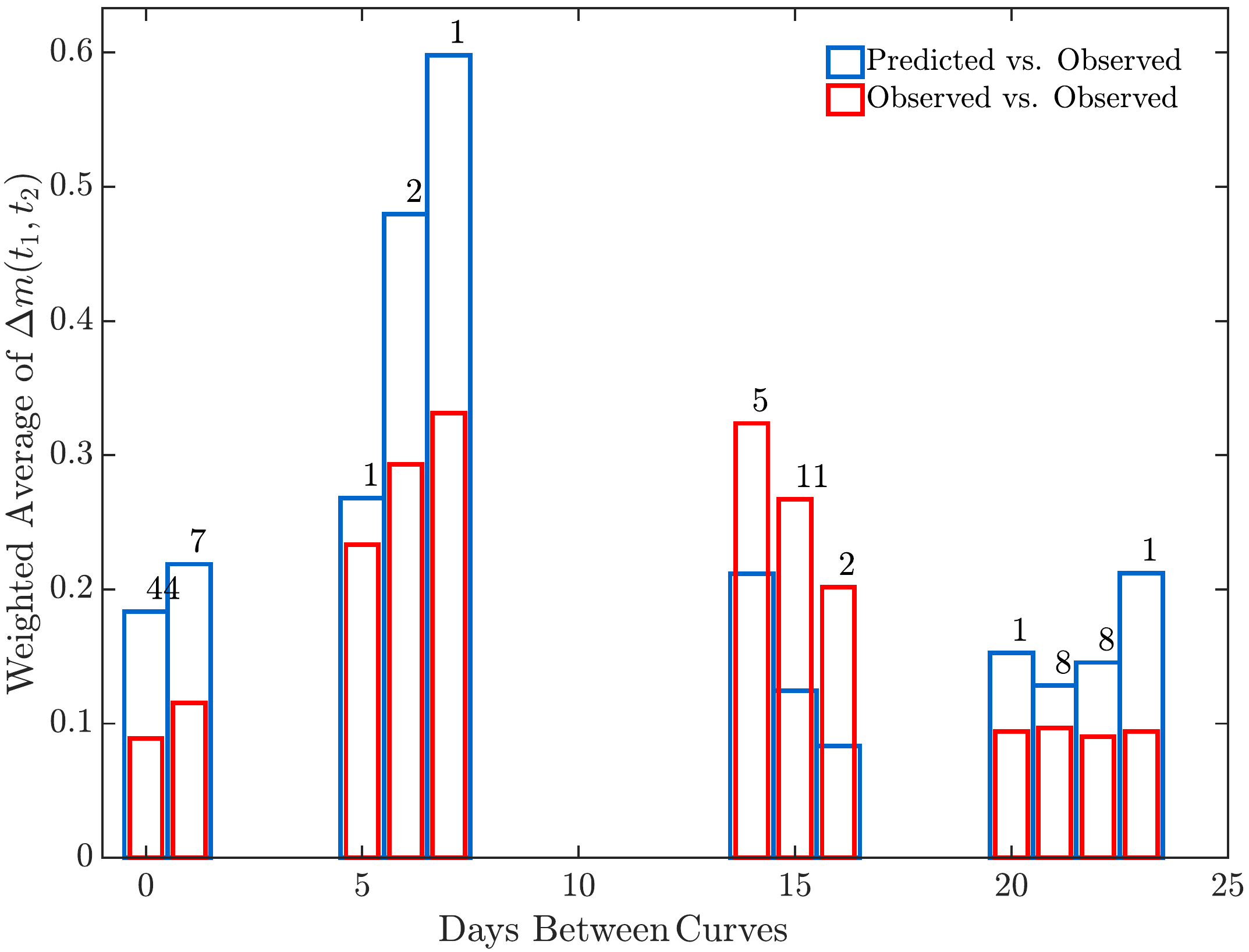}
\caption{ Error weighted averages of the absolute magnitude discrepancy (Equation \ref{eq:binDeltam}) as shown in Figure \ref{fig:RfBarMagDiscrep}, for the seismic emission correlation detector. Numbers above each bar count discrepancy observations and exclude the time-averaged performance curves $\bar{\text{Pr}}_{D}^{\text{Pre}}$ and $\bar{\text{Pr}}_{D}^{\text{Obs}}$.}
 \label{fig:SeismicBarMagDiscrep}
\end{figure}

Despite such isolated, poor estimates, the error weighted time-averages of the observed $\bar{\text{Pr}}_{D}^{\text{Obs}}$ and predicted $\bar{\text{Pr}}_{D}^{\text{Pre}}$ performance  curves generally show low magnitude discrepancies, when compared to those that quantify the STA/LTA detector. Figure \ref{fig:SeismicBarMagDiscrep} bins such error-weighted magnitude discrepancies by time duration between explosion and detection forecast. The predicted-versus-observed magnitude discrepancy bins (blue bars) are shorter than those representing the STA/LTA detector, and thereby indicate that the correlation detector has a greater predictive capability. However, these measurements do not consistently exceed the same-day observed-versus-observed discrepancy values for the correlation detector (red bars). That is, the mismatch between performance curves that we constructed using same-day data records (leftmost blue bars) sometimes exceeds the mismatch shared between pairs of observed performance curves constructed from data collected on different days (rightmost red bars). We conclude that the predictive capability of our correlation detector does not generally exceed it's day-to-day variability, and that observed performance curves can occasionally forecast correlation detection rates better than theoretical curves. 

The maximum, absolute magnitude discrepancy range ($0.9\times10^{-1}$ $\le$ $R\left( \Delta m \right)$ $\le$ $0.6$) is comparable to the peak magnitude range between predicted curves $\bar{\text{Pr}}_{D}^{\text{Pre}}$ ($\Delta m$ $=$ $0.56$). The upper value of this magnitude discrepancy range (0.6) is largely controlled by a few observed curves that outperform the theoretical curves only within the lower portion our detection band ($\text{Pr}_{D}$ $\approx$ $0.8$); theoretical and observed curves show better agreement for higher detection rates. In particular, the time averaged predicted capability for the correlation detector remains within the detection band for over a full magnitude unit. Further, while the day-to-day predictive precision of our correlation detector is comparable to that of the STA/LTA detector, the accuracy of the time-averaged performance curve, and its precision normalized accuracy ($\Delta m_{\text{Rel}}$) exceeds the corresponding accuracy of the STA/LTA detector performance curve. 

We conclude that our seismic emission correlation detector demonstrates a fair capability to forecast true detection rates, when we parameterize the cumulative probability curves with data collected from a single day. The correlation detector shows an improved (good) capability to forecast true detection rates if we exploit weighted and averaged performance curves that we collect over several days (red, stair-cased versus solid, blue curve in Figure \ref{fig:SeismicRocs}).
\begin{table*}[ht]
\caption{Magnitude Discrepancies between $\bar{\text{Pr}}_{D}^{\text{Pre}}$ and $\bar{\text{Pr}}_{D}^{\text{Obs}}$}
\small
\centering
\begin{tabular}{l | l | l | l | l | l | l | l | l |}
\hline 
Signal, Detector & $\Delta m \times 10^{-1}$ & $\Delta m_{\text{Rel}}  \times 10^{-1}$  \\  [0.5ex]
\hline 
Radio, SNR & $0.25$ & $0.68$
\\  [1ex] 
\hline
Acoustic, STA/LTA & $6.6$ & $5.6$
\\   [1ex] 
\hline
Seismic, Correlation & $1.0$ & $1.8$
\\  [1ex]
\hline
\end{tabular}
\label{table:magDiscrep}
\end{table*}
\section*{Summary and Discussion}
Table \ref{table:magDiscrep} summarizes the absolute and relative magnitude discrepancy for each detector. These data show that the time-averaged, predicted performance $\bar{\text{Pr}}_{D}^{\text{Pre}}$ of the radio emission SNR detector exceeds the time-averaged predictive performance of the seismic emission correlation detector, and that the correlation detector exceeds the time-averaged predictive performance of the acoustic emission STA/LTA detector. We conclude that SNR detector demonstrates a very good, time averaged predictive capability, while the correlation and STA/LTA detectors respectively show good and fair predictive capability.

The relative estimates of predictive performance of each signature do not reflect their mean detection capability. That is, a poor match between observed and predicted performance curves (magnitude discrepancy) does not indicate that a detector will detect the waveform recording a ``small'' explosion source with low probability. It only indicates that we have a high uncertainty in the anticipated probability that our detector will trigger on a waveform produced by that small explosion.  Specifically, Figure \ref{fig:RfRocs} indicates that a radio emission SNR detector maintains an acceptable, observed detection rate ($\text{Pr}_{D}^{\text{Obs}}$ $\ge$ $0.8$) for sources with a relative source magnitude of $m-m_{0}$ $\ge$ -0.4. The SNR detector thereby shows the lowest mean detection capability. In contrast, Figure \ref{fig:AcousticRocs} shows that an acoustic emission STA/LTA detector maintains the same detection rate for smaller magnitude sources, $m-m_{0}$ $\ge$ -0.55, and Figure \ref{fig:SeismicRocs} shows that a seismic emission correlation detector maintains that same detection rate for even smaller sources, $m-m_{0}$ $\ge$ -1.15. Therefore, a seismic emission correlation detector can target smaller sources but is less predictive than the SNR detector. We reemphasize that our objective here was not to quantify the capability of a detector to identify small explosions. Rather, our goal was to test our ability to forecast the probability with which each detector will identify an explosion triggered waveform, and to quantify our uncertainty in this forecast. 

\begin{figure}
\centering
\includegraphics[width=0.85 \textwidth]{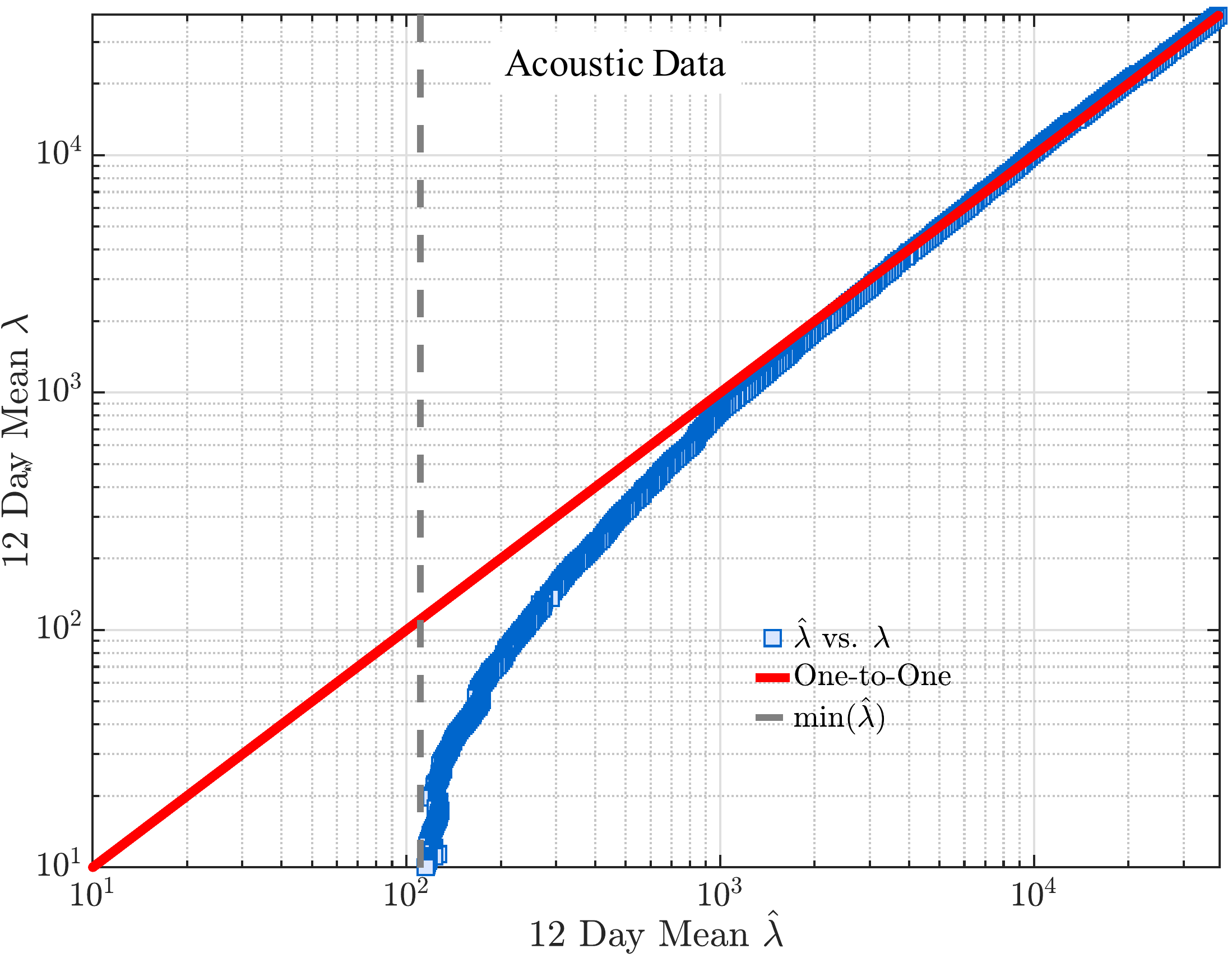}
\caption{ A log-log comparison between true values of noncentrality parameter $\lambda$ and average estimates $\hat{\lambda}$ (blue, rectangular markers) for the acoustic emission STA/LTA detector. The vertical, gray dashed line shows the minimum estimated value of $\hat{\lambda}$. The divergence of our estimates for $\hat{\lambda}$ from the red, one-to-one line indicates a positive bias at low magnitude values, where $\hat{\lambda}$ $\ge$  $\lambda$, and $\text{Pr}_{D}^{\text{Pre}}$ over-predicts true detection rates.}
 \label{fig:staLtaLambdaEstim}
\end{figure}

In quantifying such uncertainty, we note that the predicted performance of our digital detectors often exceeded their observed performance. Such mismatch can occur because of bias in our estimates of shaping parameters (like $\hat{N}_{E}$, $\hat{\lambda}$), because we permit multiple detection opportunities per processing window (in contradiction to each detector's probability model), because our we only averaged over 12 days, or through a combination of effects. We now separately consider each source of uncertainty.
To begin, we searched for bias in the PDF parameters that we estimated from each noisy data set. Each null PDF ($\mathcal{H}_{0}$ true) included several such parameters ($N_{E}$, $c$, $N_{1}$, $N_{2}$). These scalars also shaped the alternative PDFs ($\mathcal{H}_{1}$ true) that we parameterized with noncentrality parameter $\lambda$. We first checked for any bias in our estimators of the null PDF parameters. This review demonstrated that most of our estimates matched their true values well, on average. Next, we assessed our calculations of $\hat{\lambda}$. During this assessment, we empirically discovered a positive bias in Equation \ref{eq:staLtaNcEst} that over-estimated $\lambda$ for the acoustic emission STA/LTA detector, particularly for low magnitudes (Figure \ref{fig:staLtaLambdaEstim}). To understand this bias, we reviewed histogram fits between $ f_{Z} \left(  z ; \mathcal{H}_{0}  \right)$ and $\text{Hist}\vert_{2.5}^{95}(z)$. This review suggests that the degree of freedom parameter $N_{2}$ that defines the short-term processing window of the STA/LTA detector is often over-estimated for small signal amplitudes, and inflates $\hat{\lambda}$. This over-estimation occurs when the data histogram numerically fits a central $F$-PDF to data that contains some noncentral $F$ statistic samples, above its 2.5\% sample quantile (left tail). The data histogram's left-most bins therefore include some noncentral $F$-statistic samples. In these cases, our parameter estimation scheme fits a central $F$-PDF with an overestimated $N_{2}$ parameter to accommodate small $\lambda$ values. This overestimation occurs because the left tail of a corresponding noncentral $F$-PDF, with degree of freedom parameters $N_{1}$ and $N_{2}$, approximates the associated central $F$ PDF with degree of freedom parameters $N_{1}$ and $N_{2}$ $+$ $\Delta N$ ($\Delta N$ $>$ $0$).  While this bias likely inflated our predicted performance curves for the acoustic emission STA/LTA detector, we otherwise consider the mismatch between SNR and correlation detector performance curves as unattributable to mis-specification of the competing PDFs parameters.

We next considered errors in how our algorithms count waveform detections. In this case, we concede that the observed performance curves for each detector can outperform their respective predictions in certain cases because each detector permits several detection opportunities within a given processing window. Specifically, each detector accepts the maximum detection statistic value within a multi-sample window as an ostensible waveform detection. This acceptance empirically increases detection rates. The quantitative theory that we applied to form our predictions, however, does not account for such temporally adjacent, maximum statistic values. A full quantitative theory that explains detector performance between highly correlated data samples is difficult, and likely requires full Monte Carlo simulations. However, the detailed but approximate theory that we derive here provides better insight into detector behavior than blind simulations, because we can express each detector's common noncentrality parameter using relative source magnitude. This parameterization thereby affords a physical interpretation of predicted detector performance.

Third, we acknowledge that our limited data set of 12 days may not adequately represent our population of observed performance curves. More specifically, we estimated the shaping scalars for each signature's PDF from noisy data. Therefore, each performance curve is itself indexed by at least one random variable. In the case of the acoustic emission STA/LTA detector, these curves are shaped by a combination of three scalar parameters (Equation \ref{eq:staLtaParamFit}) and $\lambda$, which is a function of these parameters. Our time averaged predicted curves $\bar{\text{Pr}}_{D}^{\text{Pre}}$ therefore represent a sample mean over many realizations of such parameterized curves. We propose that we can better express our predicted curves as a marginal PDF $\mathbb{E}\{ \text{Pr}_{D}^{\text{Pre}} \}$ of $\text{Pr}_{D}^{\text{Pre}}$, given sufficient data to estimate the joint-PDF of the shaping parameters. Carmichael and Hartse (\citep{Carmichael20162}) demonstrated a similar approach for calculating performance curves and detector thresholds for a seismic correlation detector with statistic $r$ by approximating its exact PDF $f_{R}\left(r ; \mathcal{H}_{1} \right)$ with a normalized histogram. 

There are other practical caveats to our results. For instance, the source-to-receiver distance that separated the seismo-acoustic receivers ($\sim$2 km) from the explosion greatly exceeded the distance separating the radio antenna from the explosion (117 m). This distance disparity is somewhat misleading, however. We suggest instead comparing source to receiver distance in units of signature wavelength $\Lambda$.  To estimate the source-to-receiver distance in wavelength units, we first compute each signature's phase velocity $c_{p}$ by measuring first waveform arrivals in our mechanical waveform data, and taking the speed of light in a vacuum for our radio data. We then compute the dominant frequency $f_{D}$ as the geometric mean $f_{D}$ $=$ $\sqrt{f_{L}\,f_{H}}$ of our filter passband limits ($f_{L}$, $f_{H}$) for each signature. The ratio of these two terms $\Lambda$ $=$ $c_{p} / f_{D}$ estimates wavelength. These simple calculations show that the radio antennae measured electric field emissions $21$ wavelengths from its source, the acoustic receiver measured waveforms $53$ wavelengths from its source, and the seismometer measured ground motion  $18$ wavelengths from its source. Therefore, each receiver sampled its signature at the same order-of-magnitude wavelength distance from the explosion, with the acoustic receiver most remote. We concede that even this wavelength comparison is somewhat misleading, because the propagation path for each signature encounters different attenuation and scattering mechanisms. The empirical comparison in relative signal field strengths depends more on such propagation and coupling conditions than deployment distance, and such study is not within the scope of this report.

Finally, Equation \ref{eq:magAmpl} assumes a common amplitude versus magnitude relationship that may be inapplicable for substantial differences in magnitude. That is, two explosions of different source magnitudes generally emit waveforms with dissimilar frequency content. This means that their waveforms are not amplitude-scaled copies of each other, because smaller sources tend to produce higher-frequency content. Our assumption of amplitude scaling is not particular crucial for either the SNR or STA/LTA detector, which identify signals in background noise with detection statistics that are not sensitive to frequency content or waveform shape. However, our correlation detector operates under competing hypothesis tests do assume waveform amplitude scaling (Equation \ref{eq:magAmpl}). Our correlation detector results are therefore the least representative of true source scaling with magnitude, but are representative of SNR scaling.
\section*{Conclusions}
Near-ground explosions emit multiple signatures that can cumulatively offer greater evidence of explosive activity than offered by any signature alone. A goal of multiPEM research is to predictively fuse these data to better forecast how multi-signature monitoring improves detection rates of small magnitude sources. Reaching such a goal first requires the upstream capability to forecast the detection probability of each constituent signal detector, in real noise environments. Such predictive detection also requires defensible estimates of uncertainty, that preferably refer to an attribute of the explosion source (like magnitude).

To progress toward this goal, we quantify the predictive performance of waveform detectors that process radio, acoustic, and seismic signals. We defined such predictive performance as the maximum range in magnitude $\Delta m$ over which two different detection curves report the same probability of detection $\text{Pr}_{D}$, when one performance curve serves as a reference for another (Equation \ref{eq:absmagDiscrep}). This absolute magnitude discrepancy measured our detector's predictive accuracy, whereas our development of a relative magnitude discrepancy ($\Delta m_{\text{Ref}}$) measured each detector's accuracy against its predictive precision (Equation \ref{eq:magDiscrepRel}). We applied these two definitions to quantify solutions to three basic questions regarding theoretical and observed performance of waveform detectors that monitor aboveground explosions from local distances. Specifically, we (1) measured the mismatch between time-averaged, predicted and observed performance curves; (2) compared the mismatch between observed and predicted curves assembled from the same day, against the mismatch between performance curves assembled from separate days; and (3) estimated the total range in magnitude discrepancy between predicted versus observed performance curves, that we binned and averaged by day. Our data show that our radio emission detector provided the highest predictive performance ($\le 7\%$ uncertainty), followed by the seismic emission correlation detector ($\le 18\%$ uncertainty), and acoustic emission STA/LTA detector ($\le 56\%$ uncertainty). These estimates bounded the expected range in our detector's precision-normalized accuracy. Stated differently, they measured average worst-case discrepancies in our detector's predictive performance. 

Our data also show that time-averaged, predicted and observed performance curves generally match better than daily comparisons of performance curve pairs that show larger magnitude discrepancies. This same data show that our predicted curves are not noticeably biased within the interval of moderate detection probability ($0.8$ $\le$ $\text{Pr}_{D}$ $\le$ $0.98$). However, our results also show that observed performance curves measured on a particular day can sometimes provide better forecasting capability for other observed performance curves measured on a different day, when compared to theoretical performance curves calculated on that same day. This has practical consequences for several scenarios, such as monitoring missions that deploy heterogenous, instrument networks that are punctuated by interruptions in uptime. In these cases, analysts must calculate performance curves for each detector by repeatedly scaling and infusing waveforms that record the explosion into time-limited noise records (as done in Section \ref{app:Radio}, for instance). Such time-limitations correspondingly restrict analyst ability to average curves. Our results suggest than an observed performance curve that is conditioned upon limited data can forecast future detector capability often as well as a theoretical curve that is collected from the anticipated, future time period. 

We have not treated fusing the waveform evidence of a particular source using multiple signatures. However, we are currently pursuing parallel work that does accumulate such multi-signature data using Fisher's combined probability test. This work will better quantify uncertainties like magnitude discrepancy, so researchers may better bound their predictive capabilities for multiple signatures.
\section*{Acknowledgments}
This manuscript has been authored by Los Alamos National Security with the U.S. Department of Energy, under LA-UR-18-28397. The United States Government retains and the publisher, by accepting the article for publication, acknowledges that the United States Government retains a non-exclusive, paid-up, irrevocable, world-wide license to publish or reproduce the published form of this manuscript, or allow others to do so, for United States Government purposes.


\clearpage

\begin{appendices}

\numberwithin{equation}{section}
    
\section{The Acoustic Emission STA/LTA Detector} \label{sec:AcousticSTALTA}
We measured acoustic data using a mechanical differential pressure transducer deployed $\sim$2 km from the Minie shot pit that was collocated with a seismic sensor. An RT-130 digitizer sampled these pressure records at 1000 s$^{-1}$ec and continuously logged data to a hard drive throughout the testing campaign. To screen the shot-triggered waveforms from any acoustic background emissions, we developed a short-term to long-term average (STA/LTA) acoustic power detector that evaluated a binary hypothesis similar to that invoked by the radio emission SNR detector. In contrast to the radio emission data, acoustic noise power was highly variable and included high-frequency, spurious signals from clutter and long-period fluctuations in background pressure. We removed some of these features prior to applying our detector by detrending and demeaning the data, bandpassing the results between 4-20 Hz with a minimum phase, four-pole Butterworth filter, and then tapering data ends to remove oscillatory effects induced by the filter's response function. Our STA/LTA detector estimates the variance of these pre-processed data in two statistically independent, non-overlapping windows. The first, longer time window estimates acoustic noise variance $\hat{\sigma}_{l}^{2}( t < t_{S} )$ at sample $l$, over $N_{1}$ consecutive samples preceding sample $l$ ($l > N_{1} + 2$):
\begin{equation}
\label{eq:acousticVarLta}
\begin{split}
\hat{\sigma}_{l}^{2}( t < t_{S} ) &= \cfrac{1}{N_{1}} \displaystyle \sum_{k \,= \, l - 1 - N_{1}}^{l - 1} x^{2} \left( k\,\Delta t \right) \,\, \text{(LTA)}
\end{split}
\end{equation}
where we specify time $t=t_{S}$ as a dummy time index that separates the last short-time window sample from the first long-time window sample. A leading, shorter time window estimates $\hat{\sigma}_{l}^{2}( t > t_{S} )$ at sample $l$, over $N_{2}$ consecutive samples proceeding sample $l$ :
\begin{equation}
\label{eq:acousticVarSta}
\begin{split}
\hat{\sigma}_{l}^{2}( t > t_{S} )&= \cfrac{1}{N_{2}} \displaystyle \sum_{k \,= \, l + 1}^{l + 1 + N_{2}} x^{2} \left( k\,\Delta t \right) \,\,  \text{(STA)}
\end{split}
\end{equation}
Therefore, data recording an $N_{2}$ sample length signal preceded by $N_{1}$ samples of noise generated a larger STA/LTA statistic than any commensurate record containing only noise of the same variance. For Gaussian noise, such detector statistics have central $F$-distributions. To test the distributional form of these data and thereby assess the predictive capability of our detector, we performed binning experiments analogous to those we applied to the radio emission records. Specifically, we formed normalized histograms of the post-processed data over multiple duration time windows and days. These experiments demonstrated that noise data were characterized by stationary, Gaussian statistics over temporal durations of $\sim$15 min. These histograms occasionally provided a better match with mixture models that included a superposition of two Gaussian density functions. Fortunately, our tests indicated that the $\sim95\%$ left-tail of the STA/LTA acoustic detector statistic fit a central $F$-distribution even when data had such mixture model statistics, or included signal clutter. To continually assess these statistics, we included a distributional-fit error estimation scheme within our detector. This scheme computed RMS misfits between the STA/LTA statistic's normalized histogram and the best-fit central-$F$ probability density function and provided a consistent quality-check between the acoustic and radio emission detectors.

Having verified the processed acoustic data were sufficiently Gaussian, we state our competing signal models as a binary hypothesis test on two distinct distribution functions for the STA/LTA statistic. As with the SNR detector, the null hypothesis $\mathcal{H}_{0}$ presumes that data $\boldsymbol{x}$ include only noise $\boldsymbol{n}$. The alternative hypothesis $\mathcal{H}_{1}$ presumes the data $\boldsymbol{x}$ include an unknown signal $\boldsymbol{u}$ superimposed with noise. This test at sample $l$ thereby takes the form:
\begin{equation}
\label{eq:binTestStaLta}
\begin{split}
\mathcal{H}_0: &\,\, \cfrac{\hat{\sigma}_{l}^{2}(t > t_{S})}{ \hat{\sigma}_{l}^{2}(t < t_{S})} \sim c \mathcal{F}_{N_{1},N_{2}} \left( 0 \right), \,\, c, N_{1}, N_{2} \,\, \text{unknown}
\\
\mathcal{H}_1: &\,\,  \cfrac{\hat{\sigma}_{l}^{2}(t > t_{S})} {\hat{\sigma}_{l}^{2}(t < t_{S})  } \sim c \mathcal{F}_{N_{1},N_{2}} \left( \lambda_{l} \right),  \,\, c, N_{1},N_{2},\,\lambda_{l} \,\,\text{unknown}
\end{split}
\end{equation}
where $c \mathcal{F}_{N_{1},N_{2}} \left( \lambda_{l} \right)$ describes a scaled, noncentral $F$ distribution with $N_{1}$ and $N_{2}$ degrees of freedom, noncentrality parameter $\lambda_{l}$, and scaling parameter $c$.  Our STA/LTA acoustic emission detector compares these variance ratios under each hypothesis and forms a test statistic $z_{l}\left( \boldsymbol{x} \right)$ at time sample index $l$:
\begin{equation}
\label{eq:binDecRuleStaLta}
z_{l} \left( \boldsymbol{x} \right)  \triangleq \cfrac{\hat{\sigma}_{l}^{2}(t > t_{S})}{ \hat{\sigma}_{l}^{2}(t < t_{S})}   \,
\underset{\mathcal{H}_{0}}
{ \overset{\mathcal{H}_{1}}
{\gtrless}}
\,
\eta_{l}
\end{equation}
As with the SNR statistic, $\lambda$ (sample $l$ omitted hereon in this section) is a scalar proportional to the SNR and is consequently zero when data include only noise; it also quantifies the signal-versus-noise discrimination power of the hypothesis test. The asymmetry of these variance estimates (Equation \ref{eq:acousticVarLta} versus Equation \ref{eq:acousticVarSta}) changes the algebraic form of the noncentrality parameter $\lambda$, relative to that of the radio emission SNR detector. In the present case:
\begin{equation}
\label{eq:staLtaNc}
\begin{split}
\lambda &=  z(\boldsymbol{x} ) \left( \cfrac{N_{1}} {N_{2}} \right) \left( N_{2} - 2 \right) - N_{1}
\\
 &=  \text{SNR}^{}(\boldsymbol{x} ) \left( \cfrac{N_{2}} {N_{1}} \right) \left( N_{2} - 2 \right) - N_{1},
\end{split}
\end{equation}
where $\text{SNR}(\boldsymbol{x} )$ is measured over $N_{2}$-samples of the short-term window. 

We write the PDF for a random variable $s$ with a noncentral $F$ distribution under hypothesis $\mathcal{H}_j$ as $f_S\left( s; \, \mathcal{H}_j \right)$, where $\lambda=0$ under $\mathcal{H}_0$ and $\lambda>0$ under $\mathcal{H}_1$. The PDF $f_{Z_{}} \left( z; \, \mathcal{H}_j \right)$ for $z\left( \boldsymbol{x} \right)$ is therefore, by variable transformation:
\begin{equation}
\begin{split}
f_{Z_{}} \left( z; \, \mathcal{H}_j \right) &= \cfrac{1}{c} \, f_S\left( \frac{z}{c}; \, \mathcal{H}_j \right)
\end{split}
\label{eq:staLtaFPdf}
\end{equation}
%
We evaluate the screening capability of the acoustic detector for some $\lambda$ with a process analogous to that implemented for the SNR detector. As before, the threshold $\eta$ that establishes a fixed right tail probability $\text{Pr}_{FA}$ under $\mathcal{H}_{0}$ is:
 \begin{equation}
 \label{eq:probFAstalta}
 \begin{split}
\text{Pr}_{FA}  &=  \int_{\eta}^{\infty} f_{Z} \left(  z  ;  \mathcal{H}_{0}  \right) dZ
\\
&= 1 -  F_{Z} \left(  \eta  ;  \mathcal{H}_{0}  \right)
\end{split}
 \end{equation}
where $f_{Z} \left( z  ;  \mathcal{H}_{0}  \right)$ has $N_{1}$ and $N_{2}$ degrees of freedom, and $F_{Z} \left( z  ;  \mathcal{H}_{0}  \right)$ is the CDF corresponding to $f_{Z} \left(  z  ;  \mathcal{H}_{0}  \right)$. The probability of correctly identifying an acoustic signal buried in noise is the right tail probability under  $\mathcal{H}_{1}$:
 \begin{equation}
 \label{eq:probDstaLta}
  \begin{split}
\text{Pr}_{D}  &=  \int_{\eta}^{\infty} f_{Z} \left(  z  ;  \mathcal{H}_{1}  \right) dE
\\
&= 1 -  F_{Z} \left(  \eta   ;  \mathcal{H}_{1}  \right).
\end{split}
 \end{equation}
In this case, the CDF under $\mathcal{H}_{1}$ is a noncentral $F$-distributions with $N_{1}$ and $N_{2}$ degrees of freedom and (unknown) noncentrality parameter $\lambda$. We emphasize that $c$, $N_{1}$, and $N_{2}$ are also generally unknown, and estimated from data (done below). This means our detector threshold $\eta$ is also only estimable as $\hat{\eta}$ from Equation \ref{eq:probDstaLta} and is therefore a random variable. 

During our initial detector tests, we noted that degree-of-freedom parameters $N_{1}$ and $N_{2}$ that shape the PDF curves were often reduced from their theoretical values (twice the data time bandwidth product in the long and short windows). We attribute this reduction to data inter-dependency to the presence of structured noise and/or our pre-processing, band-limiting operations that introduce statistical correlation between samples. Our STA/LTA detector therefore treats these parameters as unknown scalars and estimates them in each processing window (usually limited to 15 min by noise stationarity). In our estimation process, the detector applies several degree-of-freedom estimators (numerical curve fitting strategies) and selects the resultant estimates that best fit the data-computed, normalized histogram for $Z$. Two particular estimation strategies are:
\begin{equation}
\begin{split}
\label{eq:staLtaParamFit}
\hat{c}, \hat{N}_{1}, \hat{N}_{2} &= \underset{c, N_{1}, N_{2}}{\text{argmax}} \bigr \lvert  \bigr \lvert \text{Hist}\vert_{2.5}^{95}(z) - c f_{Z} \left(  z ; \mathcal{H}_{0}  \right) \bigr \rvert  \bigr \rvert
\\
\hat{c}, \hat{N}_{1}, \hat{N}_{2} &= \underset{c, N_{1}, N_{2}}{\text{argmax}} \bigr \lvert  \bigr \lvert \text{Hist}\vert_{2.5}^{95}(z) - c f_{Z} \left( \cfrac{N_{1}}{N_{2}} z ; \mathcal{H}_{0}  \right) \bigr \rvert  \bigr \rvert
\end{split}
\end{equation}
We then selected the parameter triplet ($\hat{c}$, $\hat{N}_{1}$, $\hat{N}_{2}$) that provided the lowest fit error. We formerly measure this error by substituting the respective parameter estimates into the norm functionals of Equation \ref{eq:staLtaParamFit}, and then select the minimum of $\epsilon$  and $\epsilon^{\prime}$ :
\begin{equation}
\label{eq:staLtaHistError}
\begin{split}
\epsilon &= \bigr \lvert  \bigr \lvert \text{Hist}\vert_{2.5}^{95}(z) - f_{Z} \left(  z ; \mathcal{H}_{0}  \right) \vert_{\hat{c}, \hat{N}_{1},  \hat{N}_{2}}  \bigr \rvert  \bigr \rvert
\\
\epsilon^{\prime}&= \bigr \lvert  \bigr \lvert \text{Hist}\vert_{2.5}^{95}(z) -c f_{Z} \left( \cfrac{N_{1}}{N_{2}} z ; \mathcal{H}_{0}  \right)  \vert_{\hat{c}, \hat{N}_{1},  \hat{N}_{2}}  \bigr \rvert  \bigr \rvert
\end{split}
\end{equation}
We omit writing the error-weighted ($\epsilon$ weighted), mean performance curves for the STA/LTA detector, and refer to Equation \ref{eq:meanPrD}.
\subsection*{STA/LTA Detector Operation}
When operating our acoustic waveform detector, we established several processing parameters that met practical challenges like redundant event declaration on the same waveform. We have already documented similar processes in our description of the SNR detector, and only summarize key operations. 

As with our SNR detector, single waveforms often exceeded the STA/LTA statistic's threshold estimate $\hat{\eta}$ over the entire duration of the waveform. We avoided triggering on the same signal by specifying a time interval parameter that temporarily turned triggering ``off'' after event declaration. Therefore, when a collection of consecutive samples exceeded the concurrent event declaration threshold, the detector defined the peak value within this sample collection as ``the'' detection statistic for the underlying signal. The detector then ignored subsequent detections within the off-trigger time thereafter. Unlike the SNR detector, we set this latter time duration to equate the long-term LTA window. This choice reduced biasing our variance estimates of noise on any preceding detected waveform coda. It also avoided over-parameterizing the detector with multiple time windows. As with the radio emission detector, we additionally estimated the noncentrality parameter shaping the empirical performance curves. For the current acoustic case, we exploit Equation \ref{eq:staLtaNc} and use an estimator $\hat{\lambda}$ for $\lambda$:
\begin{equation}
\label{eq:staLtaLambdaHat}
\hat{\lambda} =  z(\boldsymbol{x} ) \left(\cfrac{ \hat{N}_{1}} {\hat{N}_{2}} \right) \left( \hat{N}_{2} - 2 \right) - \hat{N}_{1} , \,\, \text{where} \,\, z(\boldsymbol{x} ) > \hat{\eta}.
\end{equation}
In Equation \ref{eq:staLtaLambdaHat}, $z(\boldsymbol{x} )$ is the maximum value of the STA/LTA statistic over the time window that it exceeds the concurrent threshold estimate $\hat{\eta}$.
\subsection*{Semi-Empirical versus Predicted Performance  Curves}
We next tested the predictive capability of the STA/LTA detector by comparing semi-empirical performance  curves against theoretically-derived performance  curves. We constructed the empirical performance  curves by repeatedly detecting a noise-contaminated acoustic waveform that recorded the same 11.6 kg charge explosion as we selected for the radio emission detector. Specifically, we scaled this acoustic signal's original amplitude  to prescribed values that we analogously sampled from a 100-point SNR/magnitude grid (see Equation \ref{eq:magAmpl}); again, waveform shapes of smaller magnitude sources do not generally follow such a spectra-preserving scaling. Regardless, we buried the resulting waveforms in recorded noise and processed the data. Unlike the radio emission records, the infused acoustic data often included significant signal clutter. Therefore, our scaled, infused waveforms occasionally superimposed with other acoustic signals that were not attributable to a known explosion. This signal interference created variability in the observed detector performance that we did not accommodate.  These events elevated our false detection counts whenever the waveforms were localized outside our short-term (STA) detector window. With these caveats, we processed our data over all SNR grid values and only counted true detections over the entire magnitude grid, where each processing routine consumed a data window that contained 42 infused waveforms. Last, we scaled these probability computations by the number of infused waveforms in each detection window.

Figure \ref{fig:AcousticRocs} compares these empirical performance  curves against predicted performance  curves. The gray staircase plots illustrate 12 days of detection counts on infused and scaled waveforms (42 per processing window). The solid black curves show 12 days of predicted cumulative probability counts. We computed these predictions from PDFs shaped by parameters that we estimate from the data (Equation \ref{eq:staLtaParamFit}). We then computed estimates of  noncentrality parameter $\lambda$ using prescribed amplitude scaling (combining Equation \ref{eq:staLtaLambdaHat} Equation \ref{eq:magAmpl}):
\begin{equation}
\label{eq:staLtaNcEst}
\begin{split}
\hat{\lambda} &=  10^{2 (m - m_{0})} \cdot \text{SNR}_{0}^{}(\boldsymbol{x}) \left( \cfrac{\hat{N}_{2}} {\hat{N}_{1}} \right) \left( \hat{N}_{2} - 2 \right) - \hat{N}_{1}.
\end{split}
\end{equation}
Here, $\text{SNR}_{0}^{}(\boldsymbol{x})$ is the $N_{2}$-sample length SNR estimate of the original, unscaled waveform that we infused into the data $\boldsymbol{x}$ prior to detection. 
The blue solid curves in Figure \ref{fig:AcousticRocs} shows the error weighted average of these predicted performance curves $\bar{\text{Pr}}_{D}^{\text{Pre}}$, and the red staircase plot shows the weighted average of the empirical performance  curves $\bar{\text{Pr}}_{D}^{\text{Obs}}$ (Equation \ref{eq:meanPrD}).  
\section{The Seismic Multichannel Correlation Detector} \label{sec:SeisCorrDetect}
We collected seismic velocity records using a three-component Trillium Compact sensor that was co-deployed with the acoustic sensor $\sim2\,$km from the Minie shot pit. The same RT130 digitizer sampled these seismic data at 1000 s$^{-1}$, and continuously logged to a hard drive throughout the testing campaign. We used these data to detect repeatable, explosion-triggered seismic signals  that originated at the shot pit with a noise-adaptive three-channel correlation detector. Such detectors operate by scanning template waveforms that record a known event against noisy data $\boldsymbol{x}$ to identify waveforms that are amplitude-scaled copies of this template; here $\boldsymbol{x}$ records $l$ through $l+N-1$ consecutive samples of noisy ground motion that are arranged in an $N\times3$ column matrix. At each sample, these detectors thereby evaluate two competing hypotheses:
\begin{equation}
\begin{split}
\mathcal{H}_0: &\quad  \boldsymbol{x}  = \boldsymbol{n}  \sim \mathcal{N}\left( \boldsymbol{0}, \, \sigma^{2}\boldsymbol{I} \right)
\quad (\sigma \text{ unknown)}
\\
\mathcal{H}_1: &\quad  \boldsymbol{x}  = \boldsymbol{n} +  A \boldsymbol{u}  \sim   \mathcal{N}\left( A \boldsymbol{u}, \, \sigma^{2}\boldsymbol{I} \right)
\quad (A, \, \sigma \text{ unknown)}
\end{split}
\label{eq:corrHypotTest}
\end{equation}
where template $\boldsymbol{u}$ (Figure \ref{fig:SeismicTemplate}) and noise $\boldsymbol{n}$ have the same column matrix dimension as $\boldsymbol{x}$. The generalized likelihood ratio test formed from these data produces a detection statistic at sample $l$ that compares these hypotheses:
\begin{equation}
\label{eq:decRuleCorrStat}
r_{l}\left(\boldsymbol{x}\right) = 
 \cfrac{ \langle \boldsymbol{ x }, \boldsymbol{u} \rangle_{F}  }{ \vert \vert  \boldsymbol{u}  \vert \vert_{F} \vert \vert   \boldsymbol{x}  \vert \vert_{F}   }
\, \,
\,
\underset{\mathcal{H}_{0}}
{ \overset{\mathcal{H}_{1}}
{\gtrless}}
\,
 \eta_{l}
\end{equation}
where $\langle \boldsymbol{ x }, \boldsymbol{u} \rangle_{F}$ $=$ $\text{tr}\left(  \boldsymbol{ x }^{\text{T}}  \boldsymbol{ u } \right)$ denotes the Frobenius inner product (sample $l$ omitted in this section, hereon). Sample correlation $r\left(\boldsymbol{x}\right)$ provides an estimate for the population correlation $\rho$ between data $\boldsymbol{x}$ and template $\boldsymbol{u}$. As with both the SNR and STA/LTA detectors we previously introduced, the waveform amplitude $A$ is related to a scalar noncentrality parameter $\lambda$ that quantifies the signal-versus-noise discrimination power of the hypothesis test. In general, the noncentrality parameter is expressible as:
\begin{equation}
\begin{split}
\lambda &=  \cfrac{  \Vert P_{\boldsymbol{U}}^{}\left( \mathbb{E} \left\{  \boldsymbol{x} \right\}  \right) \Vert ^{2}   }{\sigma^{2}} = \cfrac{\vert \vert P_{\boldsymbol{U}} \left( \boldsymbol{u} \right) \vert \vert ^{2}}{\sigma^{2}} = \cfrac{A^{2} \Vert  \boldsymbol{u} \Vert ^{2}}{\sigma^{2}}
\end{split}
\end{equation}
where the expected value and linear-projection operators commute, and $\boldsymbol{U}$ is the rank-one subspace spanned by template $\boldsymbol{u}$; in the case data exclusively consists of noise, $\lambda$ $=$ $0$. We emphasize that if $\boldsymbol{x}$ includes a target signal that is not an amplitude-scaled copy of the template, the deterministic decorrelation between the template and target data requires a different detector (see \citep{Carmichael20162,Carmichael20163}).  When data do conform to one of the two competing hypotheses, the relation between the SNR of the waveform, $\lambda$, and the population correlation imply (algebra omitted):
\begin{equation}
\begin{split}
\rho  &=    \cfrac{ \sqrt{\lambda}  }{ \sqrt{N + \lambda -1}} 
\end{split}
\end{equation}
The performance of the multichannel correlation detector is then quantified by the Pearson Product moment PDF $f_{R}\left( r;  \mathcal{H}_{j} \right)$ $(j=0,1)$: 
\begin{equation}
\label{eq:corrBetaDist}
f_{R}(r; \mathcal{H}_{j})={\frac {(N_{E}-2)\, \Gamma  (N_{E}-1)(1-\rho ^{2})^{\frac {N_{E}-1}{2}}(1-r^{2})^{\frac {N_{E}-4}{2}}}{{\sqrt {2\pi }}\,\Gamma  \left(n-{\frac {1}{2}}\right)(1-\rho r)^{N_{E}-{\frac {3}{2}}}}}  {_{2}}F_{1} \left({\frac {1}{2}},{\frac {1}{2}};{\frac {2N_{E}-1}{2}};{\frac {\rho r+1}{2}}\right)
\end{equation}
in which $_{2}F_{1} (\cdots)$ is the Gaussian hypergeometric function whose first three arguments are parameters, and whose last argument is a variable \citep{Kowalski1972_1}. This particular PDF is parameterized by the (unknown) population correlation $\rho$  (equivalently $\lambda$) and the effective number of independent samples $N_{E}$. The false alarm on noise probability $\text{Pr}_{FA}$ is:
 \begin{equation}
 \label{eq:probFACorr}
 \begin{split}
\text{Pr}_{FA}  &\triangleq  \int_{\eta}^{\infty} f_{R} \left(  r ; \mathcal{H}_{0} \right) dr
\\
&= 1 -  F_{R} \left(  \eta \; \mathcal{H}_{0}  \right)
\end{split}
 \end{equation}
where $\mathcal{H}_{0}$ implies $\lambda$, $\rho$ $=$ $0$ and $F_{R} \left(  \bullet ;  \mathcal{H}_{0}  \right)$ is the CDF for the correlation statistic $r(\boldsymbol{x})$ corresponding to PDF $f_{R} \left(  r ; \mathcal{H}_{0} \right)$. The probability that the hypothesis test in Equation \ref{eq:decRuleCorrStat} detects a noise-contaminated target signal ($\lambda$ $>$ $0$) is similarly:
 \begin{equation}
 \label{eq:probDcorr}
 \begin{split}
\text{Pr}_{D}  &=  \int_{\eta}^{\infty} f_{R} \left(  r ;  \mathcal{H}_{1}  \right) dr
\\
&= 1 -  F_{R} \left(  \eta ;  \mathcal{H}_{1}  \right)
\end{split}
 \end{equation}
Our initial detector tests demonstrated that $N_{E}$ was significantly less than its theoretical value of twice the time-bandwidth product of the target data. This observation is consistent with results from both the STA/LTA and SNR detectors. We therefore implemented an empirical estimate for $N_{E}$ to continuously update parameterizations for $f_{R} \left(  r ; \mathcal{H}_{0} \right)$. Our estimator processes a semi-empirical target data stream with the original template, and then computes the variance of the resultant time series. This target data stream includes a wave-train composed of concatenated waveform template vectors $[\boldsymbol{u},\boldsymbol{u}, \cdots,\boldsymbol{u}]^{\text{T}}$ added to commensurate, noisy target data.  The detector algorithm then computes the sample variance $\hat{\sigma}_{R}^{2}$ of the resultant correlation time series from 99.9\% of the data by excluding $0.01\%$ of the extreme left and right tails of its histogram. This provides the needed statistic to estimate $N_{E}$ as $\hat{N}_{E}$:
\begin{equation}
\label{eq:corrEffectDOF}
\hat{N}_{E} = 1 + \cfrac{\left(1+\rho^{2}\right)^{2}}{\hat{\sigma}_{R}^{2}},
\end{equation}
and thereby compute the PDF $f_{R} \left(  r ; \mathcal{H}_{0} \right)$, and estimate the detector threshold $\eta$ as $\hat{\eta}$. Again, the fit error $\epsilon$ between a quantile-bounded histogram for the correlation detector and the optimally parameterized null PDF is analogous to our estimate for SNR or STA/LTA detector fit error:
\begin{equation}
\label{eq:corrHistError}
\epsilon = \bigr \lvert  \bigr \lvert \text{Hist}\vert_{1}^{99}(r) - f_{R} \left(  r ; \mathcal{H}_{0}  \right) \vert_{\hat{N}_{E}}  \bigr \rvert  \bigr \rvert
\end{equation}
Again, we omit writing the error-weighted ($\epsilon$ weighted), mean performance curves for the STA/LTA detector, and refer to Equation \ref{eq:meanPrD}.
\subsection*{Seismic Detector Operation}
The implementation of our correlation detector includes a waveform template $\boldsymbol{u}$ (Figure \ref{fig:SeismicTemplate}) that was absent from both the SNR and STA/LTA detectors, which leads to operational differences. In particular, correlation involves scanning $\boldsymbol{u}$ against long, $M$-sample ($M\gg N$) data streams $\boldsymbol{x}$ $\in$ $\mathbb{R}^{M\times 3 }$ as follows. 

First, we zero-pad $\boldsymbol{u}_{}$  by concatenating its rows with a matrix of zeros $\boldsymbol{0}$ $\in$ $\mathbb{R}^{\left(M - N\right)\times 3 }$ to construct an ``operational'' template $\boldsymbol{u}_{M}$ $=$ $[ \boldsymbol{u} \,;\, \boldsymbol{0}]$ $\in$ $\mathbb{R}^{M\times 3 }$ to dimensionally match the data stream that is efficient for computation. We then compute $r\left(\boldsymbol{x}\right)$ at single sample-shifts of the detection window by cross correlating $\cfrac{\boldsymbol{u}_{M}}{\vert \vert \boldsymbol{u} \vert \vert_{F}^{} }$ against $\boldsymbol{x}$ in the frequency domain, and performing data stream normalization in the time domain. Finally, we fix a false alarm probability $\text{Pr}_{FA}$ $=$ $10^{-8}$  and parameterize $f_{R} \left(  r ; \mathcal{H}_{0} \right)$ with $\hat{N}_{E}$ in order to estimate $\hat{\eta}$ from Equation \ref{eq:probFACorr}. When $r(\boldsymbol{x})$ exceeds this threshold $\hat{\eta}$ over multiple consecutive samples, our detector turned triggering ``off'' after event declaration for a time-duration equal to that of the template waveform. The detector then treats the peak absolute statistic $\vert  r(\boldsymbol{x}) \vert_{\text{max}}$ within that sample collection as ``the'' detection statistic for the underlying signal.
\subsection*{Semi-Empirical versus Predicted performance  Curves}
As with both the SNR and STA/LTA detectors, we tested the predictive capability of our multichannel correlation detector by comparing semi-empirical and theoretically-derived performance  curves. We constructed empirical performance  curves by repeatedly detecting noise-contaminated seismic waveforms that recorded the same 11.6 kg charge explosion as with our prior detectors. During this process, we scaled the amplitude of each signal recorded on the three-channels to a uniform value that we sampled from a 100-point SNR/magnitude grid (see Equation \ref{eq:magAmpl}), where the noncentrality parameter associated with magnitude grid value $m - m_{0}$ is:
\begin{equation}
\label{eq:lamCorr}
\hat{\lambda} = 10^{2( m - m_{0} )} \cfrac{\Vert  \boldsymbol{u} \Vert ^{2}}{\hat{\sigma}^{2}}
\end{equation}
where $\hat{\sigma}^{2}$ is the sample estimate of noise variance. In the present case, waveform shapes of smaller magnitude sources better follow the spectra-preserving scaling signal model under $\mathcal{H}_{1}$ in Equation \ref{eq:corrHypotTest}, at least within a magnitude unit \citep{Schaff20101,Gibbons20071}. We infused these resulting waveforms in recorded noise and processed the data with our detector; each resultant processing operation consumed 24 infused waveforms per window. As with the acoustic data, these scaled and infused waveforms occasionally superimposed with non-target signals triggered by seismic sources exterior to the shot pit. This waveform interference analogously created signal clutter that elevated the variability in the observed performance of our detector. Our evaluation scheme discounted detections on such waveforms when they were temporally separated from the infused waveform by one-half or more of the template-waveform's temporal duration. After detection, we scaled the cumulative probability computations by the number of infused waveforms in each detection window. Figure \ref{fig:SeismicRocs} illustrates our comparative results using the same color and plotting scheme as shown in Figure \ref{fig:AcousticRocs} and Figure \ref{fig:RfRocs}. 
\end{appendices}


\end{document}